\documentclass[manuscript,screen]{acmart}
\AtBeginDocument{%
  }

\setcopyright{acmlicensed}
\copyrightyear{2026}
\acmYear{2026}
\acmDOI{XXXXXXX.XXXXXXX}
\usepackage{graphicx}
\usepackage{booktabs}
\usepackage{longtable}
\usepackage{adjustbox}
\usepackage{array}
\usepackage{multirow}
\usepackage{xcolor}
\usepackage{subcaption}
\usepackage{caption}
\usepackage{capt-of}
\usepackage{comment}

\usepackage{setspace}
\begin{document}

%%
%% The "title" command has an optional parameter,
%% allowing the author to define a "short title" to be used in page headers.
\title{QKD-Integrated Quantum Noise Stream Cipher: An Overview} % of Architecture, Implementation, and Security Framework }

%%
%% The "author" command and its associated commands are used to define
%% the authors and their affiliations.
%% Of note is the shared affiliation of the first two authors, and the
%% "authornote" and "authornotemark" commands
%% used to denote shared contribution to the research.
\author{Umesh Kumar Chandra}
%\authornote{Both authors contributed equally to this research.}
\email{umesh_kc@ece.iitr.ac.in}
\orcid{0009-0003-0401-1479}
%\authornotemark[1]
\affiliation{%
  \institution{Indian Institute of Technology Roorkee}
  \city{Roorkee}
 % \state{Ohio}
  \country{India}
}

\author{Avinash Kote}
%\orcid{xxx}
\affiliation{%
  \institution{Indian Institute of Technology Roorkee}
  \city{Roorkee}
 % \state{Ohio}
  \country{India}
}
%\email{xxx}

\author{Neha Pathania}
\orcid{0000-0002-3385-4761}
\affiliation{%
  \institution{Indian Institute of Technology Roorkee}
  \city{Roorkee}
 % \state{Ohio}
  \country{India}
}

\author{Ken Tanizawa}
%\orcid{xxx}
\affiliation{%
  \institution{Tamagawa University}
  \city{Tokyo}
 % \state{Ohio}
  \country{Japan}
}

\author{Fumio Futami}
%\orcid{xxx}
\affiliation{%
  \institution{Tamagawa University}
  \city{Tokyo}
 % \state{Ohio}
  \country{Japan}
}

\author{Prem Kumar}
\orcid{https://orcid.org/0000-0002-2460-1105}
\affiliation{%
  \institution{Northwestern University}
  \city{Illinois}
 % \state{Ohio}
  \country{USA}
}

\author{Sandeep Singh}
\email{sandeep.singh@ece.iitr.ac.in}
\orcid{https://orcid.org/0000-0002-8734-9832}
\affiliation{%
  \institution{Indian Institute of Technology Roorkee}
  \city{Roorkee}
  \country{India;}
  \institution{Eindhoven University of Technology}
  \city{Eindhoven}
 % \state{Ohio}
  \country{Netherlands}
}

%%
%% By default, the full list of authors will be used in the page
%% headers. Often, this list is too long, and will overlap
%% other information printed in the page headers. This command allows
%% the author to define a more concise list
%% of authors' names for this purpose.
\renewcommand{\shortauthors}{Umesh et al.}

%%
%% The abstract is a short summary of the work to be presented in the
%% article.
\begin{abstract} % 200 words max
Quantum Noise Stream Cipher (QNSC) has emerged as a physical-layer encryption technique that exploits quantum noise and non-orthogonal coherent-state modulation to secure optical communication. However, the security of QNSC relies exceedingly on the secrecy and freshness of its seed key.  
Quantum Key Distribution (QKD), on the other hand, provides information-theoretically secure key exchange rooted in the laws of quantum mechanics. 
The convergence of these two paradigms, i.e., integrated QKD-QNSC architectures, offers a compelling solution to each of their limitations.  In such integrated systems, QKD continuously supplies and refreshes the secret seed key that governs QNSC modulation. 
Thus, governing a unified security framework that couples provably secure key establishment with high-speed quantum-enhanced physical-layer encryption. 
This work presents a comprehensive review of QNSC systems, examining their operating principles, security models under various attacks, and their integration with QKD systems. We analyze the security interplay between the key generation and encryption layers and survey experimental demonstrations and architectural progress toward practical deployment. Furthermore, we identify the open challenges and future research directions that must be addressed to realize fully integrated, quantum-secured optical communication networks at a practical scale.
\end{abstract}

%%
%% The code below is generated by the tool at http://dl.acm.org/ccs.cfm.
%% Please copy and paste the code instead of the example below.

%%%%%%%%%%%%%
% \begin{CCSXML}
% <ccs2012>
%  <concept>
%   <concept_id>00000000.0000000.0000000</concept_id>
%   <concept_desc>Do Not Use This Code, Generate the Correct Terms for Your Paper</concept_desc>
%   <concept_significance>500</concept_significance>
%  </concept>
%  <concept>
%   <concept_id>00000000.00000000.00000000</concept_id>
%   <concept_desc>Do Not Use This Code, Generate the Correct Terms for Your Paper</concept_desc>
%   <concept_significance>300</concept_significance>
%  </concept>
%  <concept>
%   <concept_id>00000000.00000000.00000000</concept_id>
%   <concept_desc>Do Not Use This Code, Generate the Correct Terms for Your Paper</concept_desc>
%   <concept_significance>100</concept_significance>
%  </concept>
%  <concept>
%   <concept_id>00000000.00000000.00000000</concept_id>
%   <concept_desc>Do Not Use This Code, Generate the Correct Terms for Your Paper</concept_desc>
%   <concept_significance>100</concept_significance>
%  </concept>
% </ccs2012>
% \end{CCSXML}

% \ccsdesc[500]{Do Not Use This Code~Generate the Correct Terms for Your Paper}
% \ccsdesc[300]{Do Not Use This Code~Generate the Correct Terms for Your Paper}
% \ccsdesc{Do Not Use This Code~Generate the Correct Terms for Your Paper}
% \ccsdesc[100]{Do Not Use This Code~Generate the Correct Terms for Your Paper}

%%
%% Keywords. The author(s) should pick words that accurately describe
%% the work being presented. Separate the keywords with commas.

% \received{20 February 2007}
% \received[revised]{12 March 2009}
% \received[accepted]{5 June 2009}

%%
%% This command processes the author and affiliation and title
%% information and builds the first part of the formatted document.
\maketitle

\section{Introduction}
The security of information has always evolved in step with the technologies used to transmit it. As optical fiber communication emerged as the backbone of global data networks, supporting gigabit-to-terabit-per-second transmission rates over intercontinental distances ~\cite{Winzer2018fiber_optics}, the need to preserve that data became more pressing. Classical cryptographic schemes such as RSA ~\cite{zhou2011research} and Diffie-Hellman ~\cite{maurer2000diffie} had long served this purpose, but their security rests on computational hardness assumptions ~\cite{cheeseman1991really}. With the advent of quantum computing, this belief began to look fragile ~\cite{Brazaola_Vicario2024_QKD_survey}. This vulnerability motivated a fundamentally different approach to secure communication, one rooted not in mathematical difficulty but in the laws of physics themselves. In 1984, Bennett and Brassard proposed the BB84 protocol ~\cite{BB84}, introducing Quantum Key Distribution (QKD) as a method of establishing secret keys whose security is guaranteed by quantum mechanics rather than computational assumptions. Any attempt by an eavesdropper to intercept quantum signals inevitably disturbs them, making intrusion detectable in principle. %With optical fiber developing as the ideal medium for its deployment, what started out as a theoretical curiosity progressively developed into a practically feasible technology. Setting the stage for more ambitious implementations, 
Early experimental demonstrations proved that secure quantum keys could indeed be distributed via standard telecommunication fibers ~\cite{PhysRevA.76.052323}. 

Despite its theoretical elegance, QKD has a practical drawback: key generation rates are essentially constrained by detector efficiency and channel loss. This makes it incompatible to be the only security mechanism for the high-capacity, high-speed data streams carried by contemporary optical networks. A complementary method was required to bridge the gap between the theoretically ideal key exchange and the realistically demanding data encryption.

The solution began to take shape in the early 2000s when H.P. Yuen proposed the Y-00 protocol, also referred to as $\alpha\eta$, within the broader framework of Keyed Communication in Quantum Noise (KCQ) ~\cite{yuen2003kcq}. The central idea was remarkably simple: instead of protecting information through complex mathematical algorithms, one could conceal it within the fundamental quantum noise that cannot be eliminated or precisely predicted. While an eavesdropper is limited by fundamental quantum measurement uncertainty, legitimate users sharing a secret key in the Y-00 protocol take advantage of a dense ensemble of non-orthogonal coherent states to enable reliable signal discrimination on their end ~\cite{hirota2004quantum, corndorf2004quantum, hirota2007practical}. This physical asymmetry results from the inability to perfectly discriminate non-orthogonal quantum states. Even if an eavesdropper uses optimal heterodyne or collective measurements, distinguishing the transmitted symbols becomes increasingly difficult as the modulation multiplicity grows ~\cite{YuenNair2006Y00Security}. Consequently, the security of Y-00 is not solely derived from computational complexity but from an irreducible physical asymmetry in quantum measurement, leading to inherent resistance against ciphertext-only and known-plaintext attack strategies ~\cite{Hirota2005_Y00Design, nair2006quantum}. In this regime, the error probability approaches that of random guessing. This effect underlies the Quantum Noise Stream Cipher (QNSC). 

Alongside the theoretical developments, the practical feasibility of the Y--00 QNSC protocol was demonstrated over a fiber-optic communication system using phase modulation ~\cite{kumar2005coherent}, intensity modulation ~\cite{Futami2014Y00Experimental}, and higher-order modulation techniques ~\cite{Yoshida2014AdaptiveQAM, Yoshida2015QAM, chen2021experimental,zhang2024quantum}, bridging the gap between theory and experiment. Beyond fiber-optic demonstrations, the applicability of Y-00 encryption has also been extended to Free Space Optical (FSO) channels ~\cite{Futami2014FSO_Y00}.

Despite the advances in transmission robustness and modulation design, QNSC fundamentally relies on pre-shared secret keys for basis selection and constellation randomization. Therefore, scalable and information-theoretically secure key distribution mechanisms become essential for practical deployment in large-scale optical networks. This naturally motivates the integration of QKD with QNSC, forming a complementary architecture in which QKD provides secure key establishment, while QNSC ensures high-speed physical-layer data encryption.

The combined study of QKD and QNSC is critically important for the advancement of modern secure communication technologies. Recent works by Nair et al. ~\cite{nair2006quantum} and Mariamichael et al. ~\cite{Mariamichael2024QNSC} have made significant contributions in theoretical principles and comprehensive overviews of QNSC systems, respectively. In detail, \cite{nair2006quantum} established the theoretical foundations of quantum-noise randomized ciphers and proved that the Y--00 ($\alpha\eta$) protocol functions as a quantum random stream cipher whose security originates from coherent-state quantum noise rather than computational hardness. It provides the formal cryptographic definitions, randomness metrics, and information-theoretic security analysis against ciphertext only and known plaintext attacks.
Subsequently, \cite{Mariamichael2024QNSC} presented a comprehensive review of QNSC implementations in optical communication, covering a wide range of modulation formats such as Intensity Modulation-Direct Detection (IMDD), Quadrature Amplitude Modulation (QAM), Phase Shift Keying (PSK), and Orthogonal Frequency Division Multiplexing (OFDM), along with reported experimental demonstrations, transmission distances, and data rates. However, these surveys primarily focus on QNSC in isolation. 
In contrast, this paper bridges theory and practice by adopting a system-level perspective: we unify diverse QNSC realizations under a generalized transmission architecture, explicitly incorporating QKD-enabled key distribution. By connecting cryptographic randomness concepts, security models, and known cryptanalytic threats with physical-layer communication metrics, this work provides a design-oriented QKD-enabled QNSC-based framework for next-generation secure optical networks.

The paper is organized as follows: Section~\ref{section:2 QKD} reviews QKD, covering discrete- and continuous-variable approaches with prepare-and-measure and entanglement-based schemes. Section \ref{section:3 QNSC} discusses QNSC, including its physical principles, historical development, and modulation-based implementations. Section \ref{section:5 security} presents a security framework for QKD-QNSC systems. Section \ref{section:4 QKD+QNSC} addresses the integration of QKD with QNSC. Finally, Section \ref{section:6 challenges and future direction} discusses the challenges and future research directions for QKD-integrated QNSC systems.

\section{Overview of Quantum Key Distribution}
\label{section:2 QKD}
 Existing cryptographic infrastructures face an existential threat from emerging quantum computing platforms, capable of efficiently solving the mathematical problems underpinning contemporary secure communication ~\cite{shor1994algorithms}. Although quantum computing poses a risk to conventional encryption methods, QKD offers a fundamentally different approach rooted in physical laws rather than computational complexity~\cite{gisin2002quantum}. QKD derives its security from the principle of the no-cloning theorem and the Heisenberg uncertainty principle.
 \begin{figure*}[h!]
\centering
\includegraphics[width=0.6\textwidth]{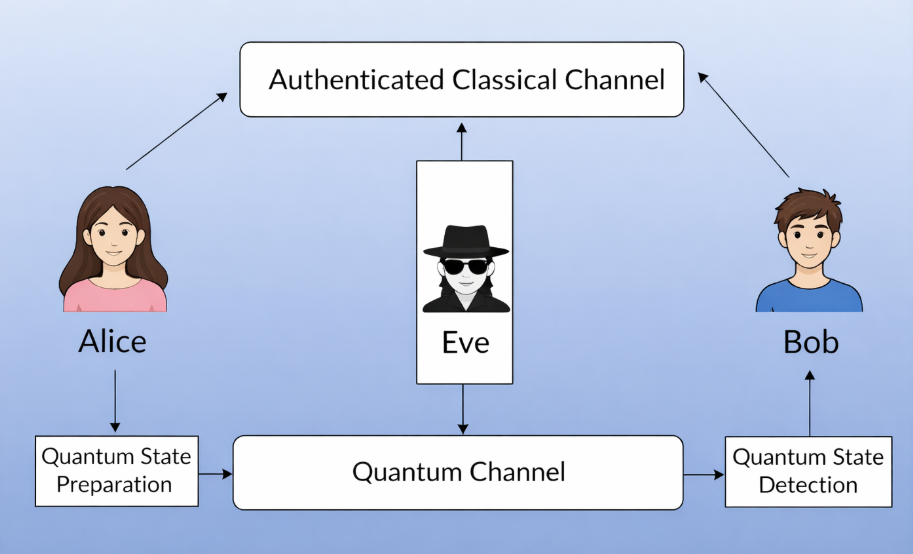}
\caption{Schematic representation of a generic QKD system illustrating quantum signal generation, transmission, detection, classical post-processing, and security monitoring.}
\Description{}
\label{qkd}
\end{figure*}

 Fig. \ref{qkd} illustrates the setup where Alice and Bob establish secure communication despite Eve attempting to intercept. Alice encodes information using non-orthogonal quantum states, which cannot be reliably distinguished through measurement. This choice is critical; if orthogonal states were used instead, Eve could measure signals in the correct basis and extract information without causing any detectable disturbance. After encoding, Alice transmits the quantum states to Bob through the quantum channel. Bob then measures the quantum signals received and examines error rates and statistical patterns in the results to estimate how much the channel has been disturbed and how much information Eve may have gained~\cite{scarani2009security}. Both parties use incompatible, non-commuting measurement bases, which guarantee that any eavesdropping attempt introduces detectable errors into the shared data.
In addition to the quantum channel, Alice and Bob rely on a public classical channel for post-processing steps such as basis reconciliation, error correction, and privacy amplification~\cite{bennett1995generalized}. Eve can freely monitor all communications on this classical channel. However, unlike the quantum channel, the classical channel must be authenticated without authentication; the protocol is exposed to man-in-the-middle attacks in which Eve impersonates one party to the other, completely breaking security \cite{stinson2005cryptography}. To guarantee unconditional security, Alice and Bob must pre-share a small secret or at least identical random strings that are partially secret \cite{wegman1981new}.

\begin{figure*}[ht!]
\centering
\includegraphics[width=1.0\textwidth]{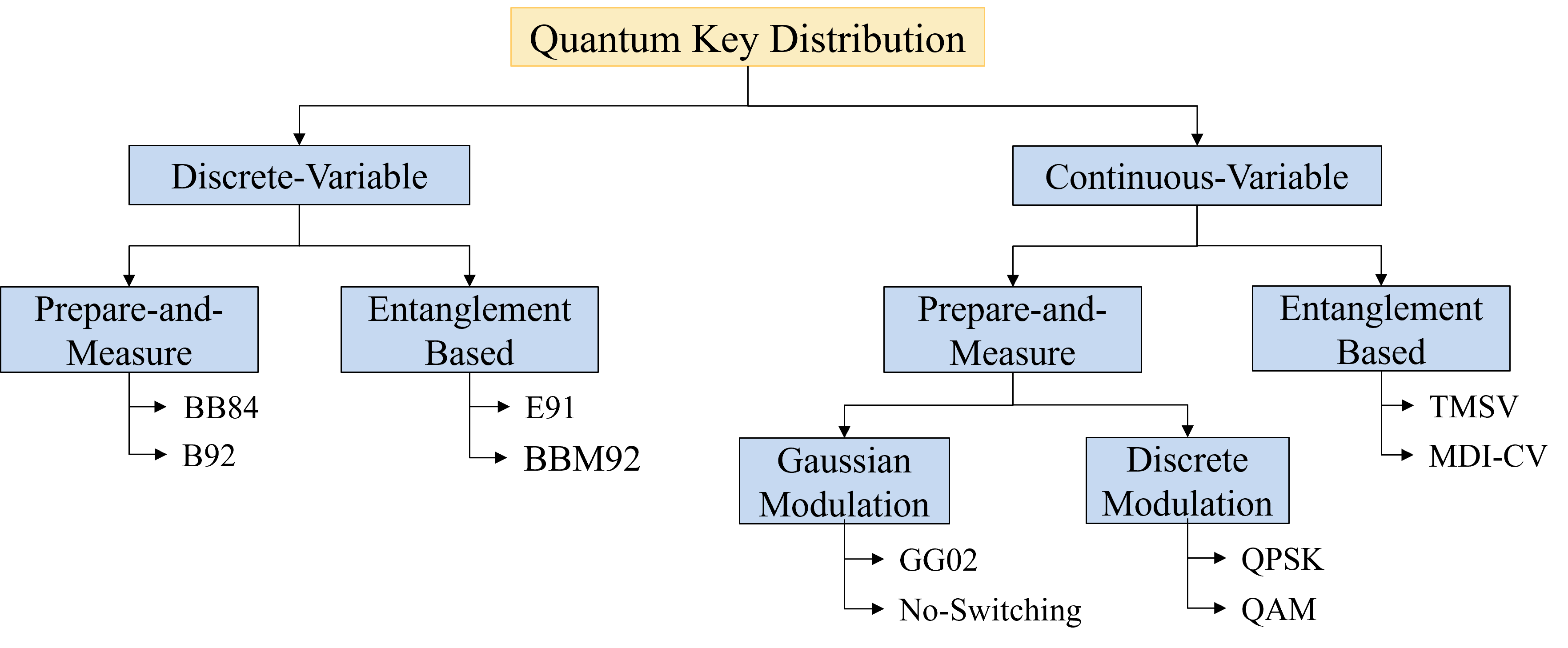}
\caption{Classification of QKD protocols based on their quantum encoding schemes.}
\Description{}
\label{qkdprotocoltypes}
\end{figure*}

Table \ref{table:qkd_papers} summarizes several proposed QKD protocols, including their key parameters and performance metrics. Additionally, Fig. \ref{qkdprotocoltypes} shows the classification of QKD protocols based on the encoding schemes: Discrete-Variable QKD (DV-QKD) and Continuous-Variable QKD (CV-QKD). Below, we briefly describe them. %In DV-QKD, information is encoded onto the polarization or phase degrees of freedom of individual photons emitted by single-photon sources~\cite{gisin2002quantum}. Photon-counting detection schemes, coupled with classical post-processing procedures, enable the extraction of secure cryptographic keys from individually detected photons ~\cite{scarani2009security}. In contrast, information encoding in CV quantum communication protocols relies on continuous electromagnetic field properties, specifically amplitude and phase quadratures ~\cite{braunstein2005quantum}. Signal quadrature components are typically extracted through homodyne detection techniques~\cite{weedbrook2012gaussian}. Fundamental security guarantees in CV-QKD stem from the Heisenberg uncertainty principle ~\cite{werner2019uncertainty}. Both of these QKD schemes are further classified into the \textit{Prepare-and-Measure (P\&M)} and \textit{Entanglement-Based (EB)} schemes. Some of the DV-QKD protocols include BB84~\cite{BB84}, Bennett-92 (B92)~\cite{B92}, Scarani-Acin-Ribordy-Gisin-04 (SARG04)~\cite{SARG04}, E91~\cite{E91} and BBM92~\cite{BBM92}. Some of the CV-QKD protocols are GG02~\cite{GG02}, Discrete Modulation (DM) QKD~\cite{leverrier2009unconditional} and entanglement-based~\cite{weedbrook2012gaussian}.

\subsection{Discrete-Variable QKD}
In DV-QKD, information is encoded onto the polarization or phase degrees of freedom of individual photons emitted by single-photon sources~\cite{gisin2002quantum}. Photon-counting detection schemes, coupled with classical post-processing procedures, enable the extraction of secure cryptographic keys from individually detected photons ~\cite{scarani2009security}. The DV-QKD is further classified as follows.
 
\subsubsection{Prepare and Measure (P\&M) Scheme}
Polarization-encoded photons, prepared by Alice, are transmitted through the quantum channel to Bob, who performs the requisite measurements upon reception~\cite{bennett1984update}. Some of the QKD protocols based on this scheme include BB84~\cite{BB84}, Bennett-92 (B92)~\cite{B92}, Scarani-Acin-Ribordy-Gisin-04 (SARG04)~\cite{SARG04}.

\subsubsection{Entanglement-Based (EB) Scheme}
Correlated photon pairs, produced by an entanglement source, are distributed to both authenticated communicants  Alice and Bob~\cite{E91}.
Projective measurements along preselected bases are then performed by Bob on his received photon. In this protocol, the quantum states of the one photon are correlated such that a measurement on one photon instantaneously determines the state of the other, independent of the spatial separation between them. Non-local correlations inherent in entangled states are
leveraged by these schemes to identify eavesdropping attempts and facilitate secure key establishment. The security of the protocol is ensured by the violation of Bell inequalities~\cite{bell1964einstein}.  In this family, QKD protocols based
on entanglement are E91~\cite{E91}, BBM92~\cite{BBM92}.

\subsection{Continuous-Variable QKD}
Information encoding in CV quantum communication protocols relies on continuous electromagnetic field properties, specifically amplitude and phase quadratures ~\cite{braunstein2005quantum}. Signal quadrature components are typically extracted through homodyne detection techniques~\cite{weedbrook2012gaussian}. Fundamental security guarantees in CV-QKD stem from the Heisenberg uncertainty principle ~\cite{werner2019uncertainty}.
 Any attempt to intercept or measure the signal inevitably disturbs its state, leaving detectable traces. Practical advantages of CV-based systems, compatibility with conventional optical infrastructure, and resilience against environmental noise position them favorably for real-world deployment~\cite{diamanti2015distributing}.
Coherent states produced by a laser source undergo amplitude and phase modulation via electro-optic modulators (EOMs), yielding Gaussian-distributed states in phase space~\cite{Grosshans2003}. The CV-QKD is further classified as follows.
\subsubsection{Prepare and Measure Scheme (P\&M)}
In the P\&M CV-QKD framework, classical random variables generated by Alice are encoded onto the quadrature components of Gaussian, coherent, or squeezed states~\cite{Grosshans2003}. Homodyne or heterodyne detection is subsequently applied by Bob to characterize the received quantum states. Correlated continuous variables obtained by both parties undergo information reconciliation and privacy amplification, ultimately distilling a secure cryptographic key~\cite{leverrier2009unconditional}. The CV-QKD P\&M protocols are further subdivided based on modulation techniques: Gaussian Modulation (GM)~\cite{GG02}, and Discrete Modulation (DM)~\cite{leverrier2009unconditional}.  
\paragraph{Gaussian Modulation (GM) Based:} 
Quadrature modulation in GM-based protocols employs Gaussian-distributed random variables, ensuring optimal resilience against collective Gaussian attacks. GG02 remains the principal exemplar of this category is the primary example of this approach~\cite{GG02}.
\paragraph{Discrete Modulation (DM) Based:}
Discrete constellation points QPSK or higher-order QAM  are utilized by Alice for information encoding in DM-based protocols ~\cite{leverrier2009unconditional}. Simplified implementation and enhanced practical performance in specific operating regimes distinguish these protocols. 
 
\subsubsection{Entanglement-Based Scheme (EB)} 
 In the EB CV-QKD framework, two-mode squeezed vacuum states, generated by the source, are distributed to Alice and Bob ~\cite{weedbrook2012gaussian}. Correlated measurement outcomes emerge when homodyne or heterodyne detection is applied to the distributed modes by both parties.

\begin{longtable}{|c|p{3.cm}|p{2.6cm}|p{1.7cm}|p{2.cm}|p{1.3cm}|p{1.5cm}|}
\caption{Overview of experimental implementations of DV,  and CV QKD protocols with their key performance parameters.}\\
\hline
\textbf{QKD Type} & \textbf{Reference (Year)} & \textbf{QKD Scheme} & \textbf{Encoding} & \textbf{Channel} & \textbf{Distance} & \textbf{Key Rate} \\
\hline
\endfirsthead
\hline
\textbf{QKD Type} & \textbf{Reference (Year)} & \textbf{QKD Scheme} & \textbf{Encoding} & \textbf{Channel} & \textbf{Distance} & \textbf{Key Rate} \\
\hline
\endhead
\hline
\endfoot
\hline
\endlastfoot

% ---------------- DV QKD (21 rows) ----------------
% \multirow{21}{*}{\textcolor{purple}{DV}}

 & Bennett et al. (1992)~\cite{bennett1992experimental} & Prepare-and-Measure & Polarisation & Free-space & 0.32 m & 10 bps \\ \cline{2-7}
 & Jennewein et al. (2000)~\cite{jennewein2000quantum} & Entanglement-based & Polarisation & Fiber & 360 m & 800 bps \\ \cline{2-7}
 & Zhao et al. (2006)~\cite{zhao2006experimental} & Prepare-and-Measure & Phase & Fiber & 60 km & 422 bps \\ \cline{2-7}
 \multirow{18}{*}[-4. em]{DV}
 & Rosenberg et al. (2007)~\cite{rosenberg2007long} & Prepare-and-Measure & Polarisation & Fiber & 102 km & 14.5 bps \\ \cline{2-7}
 & Schmitt-Manderbach et al. (2007)~\cite{schmitt2007experimental} & Prepare-and-Measure & Polarisation & Free-space & 144 km & 12.8 bps \\ \cline{2-7}
 
 & Yin et al. (2008) ~\cite{yin2007experimental}& Entanglement-based & Phase & Fiber & 123.6 km & 1.0 bps \\ \cline{2-7}
 & Peev et al. (2009)~\cite{peev2009secoqc} & Prepare-and-Measure & Phase & Fiber & 33 km & 3.1 kbps \\ \cline{2-7}
 & Liu et al. (2010)~\cite{liu2010decoy} & Prepare-and-Measure & Polarisation & Fiber & 200 km & 15.0 kbps \\ \cline{2-7}
 & Sasaki et al. (2011)~\cite{sasaki2011field} & Entanglement-based & Phase & Fiber & 45 km & 304 kbps \\ \cline{2-7}
 & Lucamarini et al. (2013)~\cite{lucamarini2013efficient} & Prepare-and-Measure & Phase & Fiber & 80 km & 120 kbps \\ \cline{2-7}
 & Tang et al. (2014)~\cite{tang2014measurement} & Prepare-and-Measure & Time-bin & Fiber & 200 km & 0.02 bps \\ \cline{2-7}
 & Wang et al. (2015)~\cite{wang2015phase} & Prepare-and-Measure & Time-bin & Fiber & 20 km & 8 bps \\ \cline{2-7}
 & Comandar et al. (2016)~\cite{comandar2016quantum} & Prepare-and-Measure & Polarisation & Fiber & 102 km & 4.6 kbps \\ \cline{2-7}
 & Liao et al. (2017)~\cite{liao2017satellite} & Prepare-and-Measure & Polarisation & Free-space & 1200 km & 1.1 kbps \\ \cline{2-7}
 & Boaron et al. (2018) ~\cite{boaron2018secure}& Entanglement-based & Time-bin & Fiber & 421 km & 6.5 bps \\ \cline{2-7}
 & Wei et al. (2019)~\cite{wei2020high} & Prepare-and-Measure & Polarisation & Fiber & 100 km & 6.2 kbps \\ \cline{2-7}
 & Yin, J. et al. (2020)~\cite{yin2020entanglement} & Entanglement-based & Polarisation & Satellite-based & 1120 km & 0.12 bps \\ \cline{2-7}
 & Pittaluga et al. (2021) ~\cite{Pittaluga2021}& Prepare-and-Measure (TF-QKD) & Phase & Fiber & 605 km & 1.1 bps \\ \cline{2-7}
 & Guarda et al (2023)~\cite{guarda2023bb84} & Prepare-and-Measure & Polarisation & Fiber & 55dB& 0.6 bps \\ \cline{2-7}
 \multirow{4}{*}[3. em]{DV}
 & Zhang, Jiawei, et al. (2025)~\cite{dou2025experimental} & Prepare-and-Measure & Polarisation & Fiber & 101.6  km & 37.6 Tbps \\
\hline

% ---------------- CV QKD (17 rows) ----------------

 & Grosshans et al. (2003)~\cite{Grosshans2003} & Prepare-and-Measure & Gaussian & Fiber & 3.1dB & 75kbps \\ \cline{2-7}
 
 & Lodewyck et al. (2007)~\cite{Lodewyck2007}{} & Prepare-and-Measure & Gaussian & Fiber & 25 km & 2 kbps \\ \cline{2-7}
 \multirow{11}{*}[-6. em]{CV}
 & Jouguet et al. (2013) ~\cite{Jouguet2013}& Prepare-and-Measure & Gaussian & Fiber & 80 km & 0.2 kbps \\ \cline{2-7}
 & Huang et al. (2016)~\cite{Huang2016OL} & Prepare-and-Measure & Gaussian & Fiber & 100 km & 0.5 kbps \\ \cline{2-7}
 & Kleis et al. (2017) ~\cite{Kleis2023}& Prepare-and-Measure & 8-PSK & Fiber & 40 km & 0.006 bps \\ \cline{2-7}
 & Zhang et al. (2019, field)~\cite{zhang2019continuous} & Prepare-and-Measure & Gaussian & Fiber & 49.85 km & 5.77 kbps \\ \cline{2-7}
 
 & Zhang et al. (2020)~\cite{Zhang2020PRL} & Prepare-and-Measure & Gaussian & Fiber & 202.81 km & 6.214 bps \\ \cline{2-7}
 & Brunner et al. (2020) ~\cite{Brunner2020}& Prepare-and-Measure & Gaussian & Fiber & 25 km & 7.04 Mbps \\ \cline{2-7}
 & Wang et al. (2022)~\cite{Wang2022PRA} & Prepare-and-Measure & QPSK & Fiber & 10 km & 133.6 Mbps \\ \cline{2-7}
 & Roumestan et al. (2022)~\cite{Roumestan2022} & Prepare-and-Measure & 256-QAM & Fiber & 50 km & 9.212 Mbps \\ \cline{2-7}
 & Wang et al. (2023)~\cite{Wang2023OL} & Prepare-and-Measure & 16-state & Fiber & 80 km & 2.11 Mbps \\ \cline{2-7}
 & Ji, Feiyu, et al. (2024) ~\cite{ji2024gbps}& Prepare-and-Measure & 	Passive-state-preparation& Fiber & 5.005 km & 1.09 Gbps \\ \cline{2-7}
 & Wu, Mingze, et al. (2025)~\cite{wu2025high} & Prepare-and-Measure  & 16 QAM & Fiber & 25 km & 18.93 Mbps 
%\hline
\label{table:qkd_papers}
\end{longtable}

\section{Quantum Noise Stream Cipher (QNSC)}
\label{section:3 QNSC}

\subsection{Foundations and Historical Development}
QNSC is a physical-layer encryption technique that improves data security by embedding encryption into the transmission process. It exploits the inherent randomness of optical noise to protect information during transmission ~\cite{yuen2003kcq}. Unlike conventional cryptographic schemes, whose security relies primarily on computational hardness assumptions, QNSC derives its security from the fundamental properties of quantum noise. The core idea of QNSC is to intentionally embed information-bearing signals within a noise background that is sufficiently large to obscure the signal structure from an unauthorized observer, while still allowing a legitimate receiver, equipped with appropriate shared secret information, to reliably recover the data.

\begin{figure}[t]
\centering
\includegraphics[width=\textwidth]{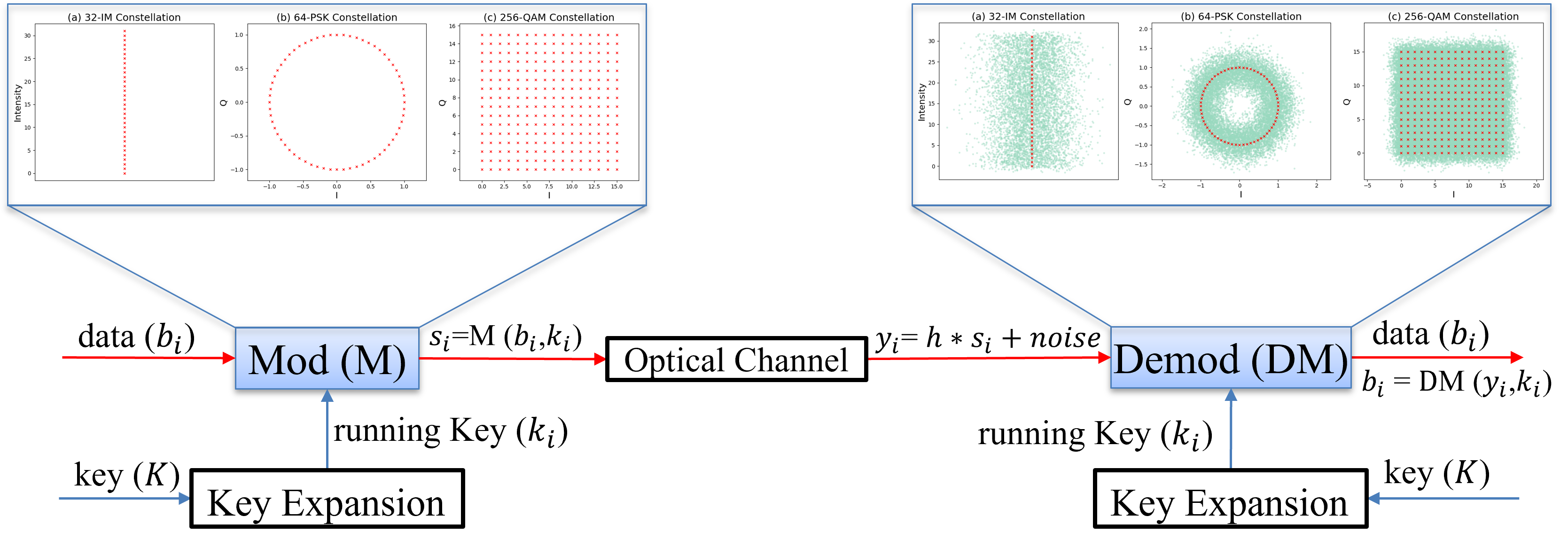}
\caption{A schematic for a QNSC-based communication system. The running key $k_i$, derived from the secret key $K$, controls the modulation process, and $h$ is the channel coefficient. Transmission over the optical channel and receiver detection noise lead to constellation spreading. 
The classical raw data and pseudo-random basis information are jointly encoded into optical signals and deliberately masked by quantum noise. The modulation block $\mathrm{Mod}\,(M)$ maps the input data bits $b_i$ onto the signal constellation, where $M$ denotes the modulation format (e.g., IM, PSK, or QAM). A running key $k_i$, generated through key expansion block, controls the modulation process. At the receiver, the block $\mathrm{Demod}\,(DM)$ performs key-assisted demodulation of the received signal $y_i$ to recover the transmitted bits $b_i$.}
\label{fig:Generalized_QNSC}
\end{figure}

The transition of Y-00 from conceptual security proposals to experimentally validated optical transmission systems accelerated significantly after 2007. High-speed intensity-modulated implementations demonstrated multi-Gbit/s secure transmission over hundreds of kilometers, confirming the feasibility of quantum-noise-masked encryption within standard IMDD fiber infrastructures ~\cite{Hirota2005_Y00Design, Hirota2007_10GbpsY00, hirota2009experiments, Harasawa2011Y00IM, Doi2010_360km}. Subsequent long-haul demonstrations extended the reach to $\sim$1000 km using multi-level intensity modulation (IM) with $M=2048$ - $4096$ basis states ~\cite{Futami2014Y00Experimental, Mao2019_LongDistanceIMDD, Futami2019_1000kmY00}, and phase level of $2^{16} - 2^{18}$   ~\cite{tanizawa2019digital20, tanizawa2019digital217, tanizawa2021ultra}, while maintaining Eve’s detection failure probability close to unity, as illustrated in Fig. ~\ref{fig:Generalized_QNSC}.

A major evolution occurred when QNSC principles were generalized from one-dimensional amplitude modulation to two-dimensional and three-dimensional coherent formats, leading to the emergence of QAM-based quantum noise stream ciphers ~\cite{Yoshida2014AdaptiveQAM, Yoshida2015QAM, chen2021experimental,zhang2024quantum}. 
More recently, the scope of QNSC has expanded beyond classical modulation randomization toward hybrid architectures integrating probabilistic shaping, multidimensional mappings, fiber-propagation-assisted encryption, and deep-learning-based end-to-end optimization ~\cite{sun2023experimental, xie2025implementation}. These advances collectively establish QNSC as a promising physical-layer cryptographic paradigm capable of providing high-speed confidentiality beyond conventional Shannon limits, while remaining compatible with modern coherent optical networks.

% -----------------------------------------------
% -----------------------------------------------

\subsection{Physical Principles of Noise-Based Security}
% Noise masking and quantum randomness...  \textcolor{blue}{implementation and block diagrams will be put here in this section}

Physical-layer security exploits fundamental limitations imposed by physics on signal observation, detection, and estimation, rather than relying solely on computational complexity. In this context, noise-based security leverages unavoidable physical noise sources such as shot noise and Amplified Spontaneous Emission (ASE) noise to limit the information that can be extracted by an unauthorized receiver. Unlike conventional cryptographic approaches, these schemes operate directly at the signal level and introduce intrinsic uncertainty into the eavesdropper measurement process ~\cite{yuen2003kcq}.

The key feature of noise-based security schemes is the creation of an asymmetric detection problem between legitimate and unauthorized receivers. The legitimate receiver possesses side information, typically in the form of a secret key through QKD, which reduces the signal discrimination task to a low-dimensional problem. In contrast, an eavesdropper lacking this information must perform blind or high-order measurements under noisy conditions, resulting in a higher error probability $(\approx 1/2)$ ~\cite{PhysRevLett.90.227901, nair2006quantum}.

\begin{figure}[t]
\centering
\includegraphics[width=\textwidth]{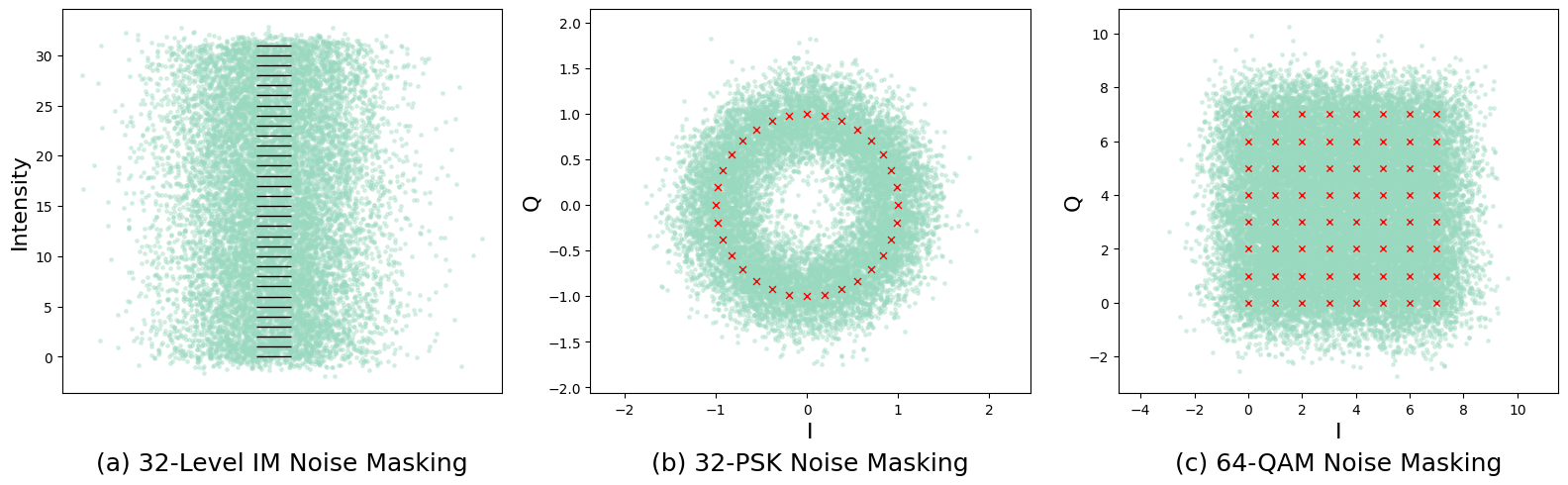} 
\caption{Illustration of noise masking in high-order modulation schemes:
(a) 32-level IM,
(b) 32-PSK constellation, and
(c) 64-QAM constellation.
Dots represent noisy received symbols, while the lines in (a) and crosses in (b, c) denote ideal constellation points.
As the modulation order increases, the spacing between adjacent signal levels  reduced, leading to significant constellation overlap due to quantum shot noise and ASE noise, effectively masking the cipher signal.}
\label{fig:constellation_noise}
\end{figure}

In optical communication systems, quantum noise plays a central role in establishing such detection asymmetry by employing noise masking. Shot noise, originating from the discrete nature of the photon, imposes a fundamental limit on amplitude and phase measurements of coherent states. Since quantum noise cannot be eliminated even in principle, it sets a lower bound on the variance of amplitude and phase estimation for coherent-state detection by an eavesdropper who does not know the correct measurement basis ~\cite{corndorf2005quantum, Futami2011ExperimentalY00}.\\
With the use of multilevel signaling or a very high modulation order, the cipher signal is masked by noise. This usage of higher modulation reduces the power difference between the adjacent levels; thus, the constellation of cipher signal is overlapped between adjacent signal levels ~\cite{futam201900} as illustrated in the Fig. \ref{fig:constellation_noise}. These noise sources include quantum noise (shot noise and ASE noise).

\subsubsection{Noise Masking by Shot Noise and ASE Noise}
Shot noise, originating from the quantum discreteness of light, represents a fundamental noise source in optical detection. In the QNSC system, this inherent randomness is intentionally leveraged to mask the transmitted signal, rather than being treated as a performance-limiting impairment.\\
In addition to shot noise, ASE noise introduced by optical amplifiers constitutes a major noise source that contributes to signal masking in fiber-based quantum noise stream cipher systems. ASE noise originates from spontaneous emission processes within optical amplifiers, such as erbium-doped fiber amplifiers (EDFAs), and is observed as intensity and phase fluctuations superimposed on the transmitted signal.\\
As a result, an eavesdropper without access to the secret running key experiences amplified uncertainty in distinguishing the signal states, while the legitimate receiver benefits from key-assisted detection. This asymmetric detection capability under identical physical channel conditions constitutes the core principle of noise masking–based security. 
The standard deviations of shot noise and ASE noise are given in ~\cite{Futami2020_Y00PhysicalLayer} as
\begin{equation}
\sigma_{\text{shot}} = e \sqrt{\frac{2 P_{\text{0}} B}{h \nu_0}}
\;\text{and}\;
\sigma_{\text{ASE}} = \sqrt{\frac{2 P_0^2 B}{R_{\mathrm{REF}}\,\mathrm{OSNR}}}.
\end{equation}

The noise masking number for different modulation schemes has been investigated in
~\cite{Futami2020_Y00PhysicalLayer, jiao2017physical, futami2019theoretical, Tanizawa2020MicrowavePSKY00, Nakazawa2014QAMQSC, Tanizawa2021OFDMQuantumNoise, Yoshida2015QAM}.

For an IMDD system ~\cite{Futami2020_Y00PhysicalLayer}, the multi-level intensity-modulated signals, denoted by $P_i$ ($1 \le i \le 2M$), are considered, where the number of intensity levels and bases are $2M$ and $M$, respectively. The amount of noise masking $(\Gamma)$ is expressed as

\begin{equation}
\Gamma_{\mathrm{IM,shot}}
= \frac{2 \sigma_{\text{shot}}}{\Delta P}
= \frac{2(2M - 1)e}{P_{2M} - P_1}
\sqrt{\frac{2 B P_0}{h \nu_0}}
\;\text{and}\;
\Gamma_{\mathrm{IM,ASE}}
= \frac{2\sigma_{\mathrm{ASE}}}{\Delta P_{\mathrm{basis}}}
= \frac{2(2M-1)P_{0}}{P_{2M}-P_{1}}
\sqrt{\frac{2B}{R_{\mathrm{REF}}\mathrm{OSNR}}}
\label{eq:Nsigma}
\end{equation}

where
\begin{equation}
\Delta P = \frac{P_{2M} - P_1}{2M - 1}.
\end{equation}

For PSK systems under shot-noise-limited conditions ~\cite{tanizawa2021ultra}, the noise masking number is given by
\begin{equation}
\Gamma_{\mathrm{PSK}} = \frac{M \cdot 2^m}{2\pi} \sqrt{\frac{2 h \nu_0 B}{\eta_q P_0}}
\label{eq:Gamma_PSK}
\end{equation}

For QAM systems under shot-noise-limited conditions, the noise masking number is expressed as
\begin{equation}
\Gamma_{\mathrm{QAM}}
= \frac{\pi \left(M \cdot 2^{2m} - 1\right) h \nu_0 B}{12 \eta_q \overline{P}_0}
\label{eq:Gamma_QAM}
\end{equation}

Here, $M$ is the modulation order, $m$ is the bit resolution of signal randomization, $h\nu_0$ is the photon energy at the optical carrier frequency, $e$ is the electron charge, $B$ is the signal bandwidth, $R_{\mathrm{REF}}$ is the reference bandwidth (typically 12~GHz (0.1~nm)), $\eta_q$ is the quantum efficiency of the photodetector, $P_0$ is the received optical power, $\overline{P}_0$ denotes its average, and $\mathrm{OSNR}$ denotes the optical signal-to-noise ratio of the Y-00 cipher signal.

\subsection{Modulation Schemes in QNSC}
The development of QNSC scheme has progressed through multiple optical transmission techniques and modulation schemes. A comprehensive comparison of these developments is provided in Table~\ref {tab:IMDD_PSK_QAM}. Initial demonstrations employ QNSC scheme (Y–00 protocol),  implemented over IMDD systems using multi–level amplitude modulation format, such as 4096–level Amplitude Shift Keying (ASK), to realize secure optical transmission over long–haul fiber links ~\cite{grigoryan2004long, Kumar2004, Eric2004, hirota2004quantum, Hirota2005_Y00Design,  nair2006quantum, Grigoryan2005_QNRDataEncryption, Hirota2007_10GbpsY00, shimizu2008running, hirota2009experiments, Kanter2009_PhysicalLayerEncryption, Doi2010_360km, Futami2014Y00100Gbps, Hirota2010BeyondShannon, Futami2011ECOCY00, Harasawa2011Y00IM, Lu2012CSIPHomodyne, Futami2014Y00Experimental, Futami2017Y00Transceiver,
futami2017experimental, Tanizawa2018CoarseFineY00, Futami2019_1000kmY00, futam201900, Futami2020_Y00PhysicalLayer, Jiao2020_100kmSecureFiber, Futami2020_SecureFSO, wang2021experimental, yu2020secure, zhu2023optical, kato2023quantum, xiao2023physical}. 

Subsequent studies extended the modulation space to include high–order M-ary PSK and QAM formats ~\cite{Eric_feb2004_SPIE, corndorf_2004_cleo, corndorf2005quantum, nair2006quantum, Banwell_2005_MILCOM,Gregory_2005_SPIE, Grigoryan2005_QNRDataEncryption, Chuang_2005_PTL, kato2008quantum, Nakazawa2014QAMQSC, Yoshida2015QAM, Yoshida2015ECOC, Yoshida2015ofc, yoshida2016single, nakazawa2016real, futami201700, tanizawa2018digital, wang2018ciphertext, tanizawa2018tradeoffs, tanizawa2019practical, tanizawa2021ultra, li2021ciphertext, li2022q, matsumoto2024high}. Advancements in PSK modulation schemes, including multi-ring and dual-polarization PSK, have expanded the effective constellation space, thereby enhancing transmission reach under quantum-noise masking ~\cite{wang2019multiring,tanizawa2019single,tanizawa2019digital20}. In parallel, QAM-based QNSC systems have been proposed, exploiting digital coherent transceivers to overlay quantum noise onto high-order QAM constellations (4–128 QAM) ~\cite{Yoshida2014AdaptiveQAM}. As a result, high-capacity secure coherent transmission over distances approaching hundreds and thousands of km has been demonstrated ~\cite{nakazawa2015qam, yoshida202110,lei202116, sun2023experimental}. Furthermore, to achieve a higher modulation order $(2^{17})$, the Coarse and fine modulation is proposed in which multilevel signaling is performed ~\cite{Tanizawa2018CoarseFineY00, tanizawa2019digital217}. QAM-based QNSC technique is more robust against eavesdroppers than IM/ASK or PSK ~\cite{Nakazawa2014QAMQSC}.\\

\begin{figure}[t]
\centering
\includegraphics[width=\textwidth]{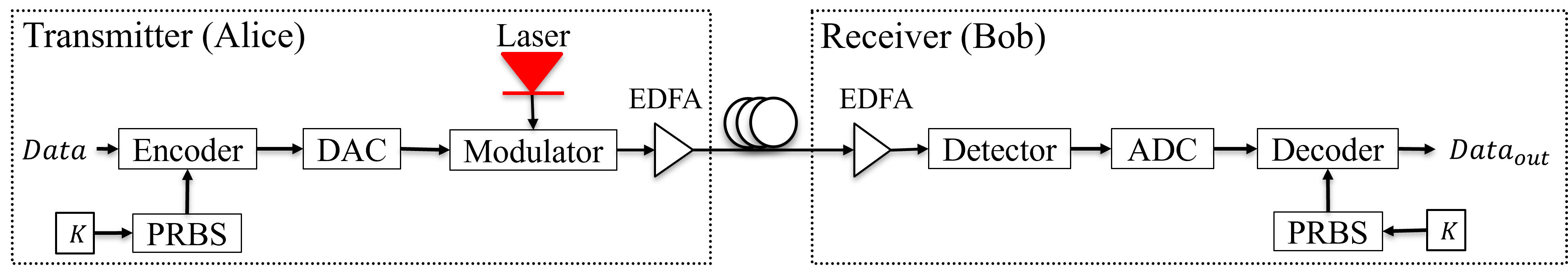}
\caption{Generalized Alice–Bob architecture of a QNSC-based secure optical communication system, illustrating PRBS-driven key-controlled physical-layer encryption at the transmitter, optical fiber transmission, photo-detection and synchronized decryption at the receiver.}
\label{fig:QNSC_schematic}
\end{figure}

\subsection{Implementation of Quantum Noise Stream Cipher (QNSC)}
Figure~\ref{fig:QNSC_schematic} illustrates the conceptual implementation of the QNSC-based optical communication framework. A shared secret key $K$ is applied to the PRBS block to control the modulation basis selection at the transmitter (Alice). For each bit or symbol, the input data is mapped onto one of the $M$ possible modulation basis states according to the secret key and the employed modulation format, such as IM, PSK, or QAM. At the receiver (Bob), the synchronized PRBS and decoder blocks use the same shared key to recover the transmitted data.

The modulated optical signal is transmitted as a coherent state through an optical fiber channel. During propagation, amplification, and reception, the signal is inherently affected by quantum noise, including shot noise at the photo-detector and ASE noise introduced by optical amplifiers. This quantum noise results in masking of the transmitted symbols in the modulation space. 
At the legitimate receiver (Bob), the same running key sequence is employed to perform key-assisted demodulation, enabling correct basis alignment prior to detection. As a result, Bob can reliably recover the transmitted data with a low bit error rate (BER). In contrast, an eavesdropper (Eve), who does not possess the secret key, must perform blind detection without knowledge of the correct modulation basis. This lack of basis information, combined with quantum noise masking, leads to a significantly increased BER at Eve.

The implementation of higher-order modulation format (QAM)-based QNSC has been experimentally demonstrated in fiber-based systems ~\cite{Yoshida2014AdaptiveQAM, Yoshida2015QAM, chen2021experimental,zhang2024quantum}. 
Additionally, Nakazawa and co-workers showed that encrypting conventional 16-QAM data into ultra-dense high-order constellations results in a quadratic scaling of the noise masking number $\Gamma$, thereby dramatically increasing brute-force complexity for an unauthorized observer ~\cite{Nakazawa2014QAMQSC, yoshida2016single}. This direction enabled record coherent QNSC demonstrations, including real-time 70-Gbit/s transmission with online key renewal via continuous-variable QKD ~\cite{nakazawa2016real, nakazawa2017qam}, and ultimately multi-terabit coherent encrypted fiber systems ~\cite{sun2024field, yoshida202110} exceeding 98.81 Tbit/s capacity with wavelength-division-multiplexed (WDM) system over 112 km.

The implementations of the QNSC in optical wireless communication (OWC) remain limited compared with fiber-based systems. FSO experimental demonstrations have shown error-free Y-00/QNSC transmission at data rates up to 2.5~Gbit/s over short-range FSO links with transmission distances on the order of millimeters to several tens of meters ~\cite{Futami2014FSO_Y00, Futami2020_SecureFSO}. In particular, secure FSO transmission using 4096-level IM has been experimentally verified over a 30~m free-space link, achieving a BER below $10^{-9}$, and link-budget analysis with optical pre-amplification indicates the feasibility of extending the transmission distance to approximately 1~km under sufficient optical signal-to-noise ratio conditions ~\cite{Futami2020_SecureFSO}. More recently, end-to-end modeling of QNSC-enabled FSO systems has been investigated for transmission distances up to 10~km under atmospheric turbulence using deep-learning-based channel modeling, focusing on waveform reconstruction accuracy and statistical indistinguishability rather than direct BER measurements ~\cite{zhang2025end}. Experimental demonstrations of underwater optical wireless communication (UOWC) have reported the successful realization of a $6m$ link, validating the feasibility of optical data transmission in an underwater environment ~\cite{shang2018underwater}.  

%Beyond fiber-optic demonstrations, the applicability of Y-00 encryption has also been extended to Free Space optical (FSO) channels ~\cite{Futami2014FSO_Y00}, where Futami and Hirota experimentally demonstrated a 2.5-Gbit/s secure FSO transmission employing 4096-level intensity-modulated Y-00 signals, targeting aviation and short-range wireless optical security scenarios ~\cite{Futami2014FSO_Y00} and later secure real-time FSO implementations employing 4096-level intensity-modulated Y-00 signals up to 30m is demonstrated ~\cite{Futami2020_SecureFSO}. Recently, Zhang et al. ~\cite{zhang2025end} proposed an end-to-end deep-learning framework to model PAM-based QNSC transmission under Gamma--Gamma atmospheric turbulence. Security was evaluated through statistical indistinguishability metrics such as probability density function(PDF) matching and Kullback-Leibler (KL) divergence, demonstrating the robustness of quantum-noise-masked ciphertext against interception in fading channels.

The security of QNSC is quantified by the noise masking number $\Gamma$, which characterizes the degree of overlap between adjacent constellation levels due to quantum noise. Larger values of $\Gamma$ correspond to stronger masking of the transmitted symbols and enhanced resistance against interception and estimation attacks. To further increase the Security of the stream cipher, various randomization/mapping methods were proposed, including irregular mapping, overlapping selection keying (OSK) ~\cite{Hirota2007_10GbpsY00}, deliberate signal randomization (DSR) ~\cite{YuenNair2006Y00Security, futami2022experimental, futami2023field}, bitwise NOT operations~\cite{wang2020theory, wang2018ciphertext}, multi-ring amplitude mapping~\cite{wang2019multiring}, and key-dependent constellation rotation~\cite{wang2019multibit} \& Gray code based mapping~\cite{li2021ciphertext}. These mappings introduce detection asymmetry between legitimate and unauthorized receivers and transform the eavesdropper’s observation into a noise-masked mixture of signal states, thereby significantly degrading blind detection performance and making the system immune to attacks ~\cite{kato2016quantum}.

In addition to mapping-based security enhancements, hybrid encryption architectures have been investigated in ~\cite{11430112}. In this scheme, the data stream is first encrypted with the Advanced Encryption Standard (AES) encryption and subsequently transmitted using Y--00-based encryption. In this scheme, AES operates at the data layer while the Y-00 system provides physical-layer security through quantum noise masking, resulting in a complementary combination of computational and physical-layer security mechanisms.

% \begin{equation}
% {\sigma}_{\mathrm{PSK/QAM}}
% =
% \sqrt{
% \frac{1}{2M^{2}}
% \sum_{n=1}^{M^{2}}
% \left(
% \sigma_{I,n}^{2}
% +
% \sigma_{Q,n}^{2}
% \right)
% }
% \end{equation}
% and the amount of noise masking $(\Gamma)$ is
% \begin{equation}
% \Gamma_{\mathrm{PSK/QAM}}
% =
% \left(
% \frac{2{\sigma}_{\mathrm{PSK/QAM}}}{\Delta}
% \right)^{2}
% \end{equation}
% where 
% \mathrm{\Delta = 2/(M-1)}

% \subsubsection{Noise Masking by Shot noise}

% \begin{equation}
% \sigma_{\text{shot}} = e \sqrt{\frac{2 \eta P_{\text{opt}} B_e}{h \nu}}
% \end{equation}

% \subsubsection{Noise Masking by ASE noise}

% \begin{equation}
% \sigma_{\text{ASE}} = \sqrt{\frac{2 P_0^2 B}{R_{\text{ref}} \, \text{OSNR}}}
% \end{equation}

\begin{longtable}
%{|p{3cm}|p{2.5cm}|p{2.5cm}|p{2cm}|p{4.5cm}|}
{|p{2.5cm}|p{2.5cm}|p{1.8cm}|p{7.2cm}|}
\caption{Experimental and Theoretical Demonstrations of IM/PSK/QAM-based Y-00/Quantum Noise Stream Cipher Systems}\\

\hline
\textbf{Reference}   &
\textbf{Modulation Type} & 
\textbf{Data Rate} & 
\textbf{Key Results} \\
\hline
\endfirsthead

\hline
\textbf{Reference} &
\textbf{Modulation Type} &
\textbf{Data Rate} &
\textbf{Key Results} \\
\hline
\endhead

\hline
\endfoot

\hline
\endlastfoot

Hirota \& Sohma ~\cite{Hirota2005_Y00Design} &
IM, M=100$\sim$200 &
1 Gbps &
20 km fiber experimental transmission under heterodyne attack, and Eve assumed unlimited computational power \\
\hline

Grigoryan et al. ~\cite{Grigoryan2005_QNRDataEncryption} &
M-ASK and M-PSK, $M=2048$ &
10 Gbps &
M-PSK$\sim$5000 km; M-ASK$\sim$720 km \\
\hline

Harasawa \& Hirota ~\cite{Harasawa2011Y00IM} &
4096-level IM ($M=2048$) &
2.5 Gbps &
192 km; BER vs received power and division number; STM-16 / OC-48 transceivers \\
\hline

Futami et al. ~\cite{Futami2020_Y00PhysicalLayer} &
4096 levels
(M = 2048) IM &
1.5 Gbps &
1000 km transmission; BER $=10^{-9}$ is 30 dB OSNR and $P_{\mathrm{in}}=5$--7 dBm; minimum and maximum detection probability $1.7\times10^{-55}$ ($\Gamma_{\mathrm{IM}}=190$) and $1.6\times10^{-38}$ ($\Gamma_{\mathrm{IM}}=330$) \\
\hline

Futami et al. ~\cite{Futami2020_SecureFSO} &
4096-level IM over FSO &
1.5 Gbps &
30 m free-space transmission; achieved BER $<10^{-9}$ at zero additional loss and with pre-amplification and 13 dB additional loss; projected $\sim$1 km FSO reach \\
\hline

Wang et al. ~\cite{wang2021experimental} &
$2^{14}$-level PAM (DD-MZM) &
100 Gbps and 40 Gbps with proposed scheme &
50 km transmission with BER below $3.8\times10^{-3}$; optical CTFM enables secure 100 Gbps; proposed optical-power-based CTFM improves modulation depth and provides $\sim$5 dB power gain over traditional \\
\hline

Zhu et al.
~\cite{zhu2023optical} &
256-level IMDD &
156.25 Mbps &
25 km transmission; dual-layer security with stealth via ASE-noise embedding and confidentiality via Y-00 basis randomization; security analyzed using number of masked symbols ($\Gamma \ge 1$) \\
\hline

% \end{longtable}

% \begin{longtable}
% %{|p{3cm}|p{2.5cm}|p{2.5cm}|p{2cm}|p{4.5cm}|}
% {|p{3cm}|p{2.5cm}|p{2.5cm}|p{6.cm}|}

% \caption{Experimental and Theoretical Demonstrations of Phase-based Y-00 / Quantum Noise Stream Cipher Systems}
% \label{tab:Phase_QNSC}

% \hline
% \textbf{Reference} &
% \textbf{Modulation Type} &
% \textbf{Data Rate} &
% \textbf{Key Results} \\
% \hline
% \endfirsthead

% \hline
% \textbf{Reference} &
% \textbf{Modulation Type} &
% \textbf{Data Rate} &
% \textbf{Key Results} \\
% \hline
% \endhead

% \hline
% \endfoot

% \hline
% \endlastfoot

Corndorf et al. ~\cite{corndorf2005quantum} &
Phase modulation  &
650 Mbps &
200 km fiber; polarization-mode demonstrates encryption efficacy;  time-mode demonstrates compatibility with existing telecom infrastructure \\
\hline

% Nair et al. 
% ~\cite{nair2006quantum} &
% M-ary coherent-state modulation (PSK / ASK / IMDD) &
%  &
% Foundational theory of quantum-noise randomized ciphers; security analyzed via error probability, guessing probability, and unicity distance; comprehensive survey of $\alpha$--$\eta$ systems \\
% \hline

% Grigoryan et al. ~\cite{Grigoryan2005_QNRDataEncryption} &
% M-ASK and M-PSK ($M=2048$) &
% 10 Gb/s &
% M-PSK transmission up to 5000 km; M-ASK limited to $\sim$720 km; comparative BER performance under different quantum noise masking conditions \\
% \hline

% Tanizawa \& Futami  ~\cite{tanizawa2018digital} &
% PSK (BPSK, Y-00 64-PSK, Y-00 1024-PSK) &
% 10 Gb/s &
% Experimental comparison shows negligible OSNR penalty due to Y-00 encryption and decryption compared with conventional BPSK \\
% \hline

% Tanizawa \& Futami ~\cite{tanizawa2018tradeoffs} &
% $2^{16}$-level PSK &  10Gb/s
%  &
% Analytical trade-off between quantum noise masking number and transmission reach; masking dominated by shot noise; reach decreases with increasing masking level \\
% \hline

% Wang et al. 
% ~\cite{wang2019multiring} &
% Multi-ring BPSK (concentric M-PSK) with OFDM &
%  &
% Enhanced security over BPSK-QNSC; running-key SER improvement of 0.22 at OSNR = 50 dB with 256 bases; Bob performance unchanged with $\sim$2 dB OSNR difference \\
% \hline

Tanizawa \& Futami ~\cite{tanizawa2019digital217} &
$2^{17}$-level PSK with coarse-to-fine modulation & 10-Gbaud  &
400 km transmission; BER below $3.8\times10^{-3}$ at OSNR $\sim$5.5 dB and launch power $\sim$-23 dBm. \\
\hline

Tanizawa et al. ~\cite{tanizawa2019single, tanizawa2021ultra} &
Dual-polarization PSK \newline ($2^{18}$ phase levels)&
48 Gbps line rate \newline (40 Gbps net) &
~\cite{tanizawa2019single} -- 400 km and 800 km transmission; OSNR penalty due to encryption/decryption is $\sim$0.5 dB at the SD-FEC threshold, and ~\cite{tanizawa2021ultra} -- Maximum reach of 8,094 km at $P_{\mathrm{in}}=-10$ dBm and 10,118 km at $P_{\mathrm{in}}=-7$ dBm using a recirculating loop; Q-penalty of $\sim$0.3--0.4 dB; Ideal heterodyne detection without the seed key yields $\mathrm{SER}_{\mathrm{eve}}\approx0.9975$.
\\
\hline

Zhu et al. ~\cite{zhu2022quantum} &
QNSC using equivalent spectral phase encoding &
10 Gbps &
Ciphered signal appears noisy in time and frequency domains; Error-free stealth transmission over $>$80 km and successful operation over $>$400 km SMF  under ASE noise masking. \\
\hline

Tanizawa et al ~\cite{tanizawa2023tradeoff} &
PSK over WDM; QPSK data encrypted to $2^{(m+2)}$-PSK &
10 Tbps &
Theoretical analysis of reach--security trade-off in nonlinear fiber channels using the GN model. Results indicate $\sim$10,000 km reach with $>$10 Tbit/s.
\newline
$m$: phase randomisation bits
\\
\hline
% \end{longtable}

% \begin{longtable}
% {|p{3cm}|p{2.5cm}|p{2.5cm}|p{6cm}|}

% \caption{Experimental and Theoretical Demonstrations of QAM-based Y-00 / Quantum Noise Stream Cipher Systems}
% \label{tab:QAM_QNSC}
% \hline
% \textbf{Reference} & \textbf{Modulation Type} & \textbf{Data Rate} & \textbf{Key Results} \\
% \hline
% \endfirsthead

% \hline
% \textbf{Reference} &
% \textbf{Modulation Type} &
% \textbf{Data Rate} &
% \textbf{Key Results} \\
% \hline
% \endhead

% \hline
% \endfoot

% \hline
% \endlastfoot

Yoshida \emph{et al.} ~\cite{Yoshida2014AdaptiveQAM} &
Polarization-multiplexed 4/16/64 QAM (adaptive) &
20--60 Gbps &
320 km transmission; FPGA-based TX/RX with OPLL; Error-free operation (BER $<10^{-11}$ with FEC); OSNR $>$22 dB (B2B) and $>$26 dB (320 km); EVM degraded to 2.6\% (160 km) and 3.6\% (320 km). OSNR margin improved by 9 dB (64$\rightarrow$16 QAM) and 17 dB (64$\rightarrow$4 QAM) \\
\hline

Nakazawa \emph{et al.} ~\cite{Nakazawa2014QAMQSC} &
16 QAM encrypted into 32$\times$32 to 4096$\times$4096 QAM &
10 Gbps &
160 km transmission; Error-free transmission (BER $<3\times10^{-5}$ at $P_{\text{out}} > -38$ dBm (16 QAM) and $>-41$ dBm (4-ASK). Strong eavesdropper DFP \\
\hline

Yoshida \emph{et al.} ~\cite{yoshida2016single} &
Polarization-multiplexed  16 QAM encrypted with 4096$\times$4096 QAM &
40 Gbps &
480 km Error-free transmission (BER $<10^{-9}$ at $P_{\text{out}}>-27.5$ dBm and BER $<2\times10^{-3}$ at $P_{\text{out}}>-36.5$ dBm); DFP reached 99.7\% at $-35$ dBm and $>99.998$\% for multiplicity $M=4096$. \\
\hline

Futami \emph{et al.} ~\cite{futami201700} &
PDM 16-QAM WDM overlaid with  IM-based Y-00 with 4096 levels &
256 Gbps &
320 km WDM transmission; Effective data rate 200 Gbps per WDM channel; Optimum launch power 0 dBm/ch, achieving Q-factor 11.4 dB: BER $ < 10^{-9}$ for launch powers $>2$ dBm (OSNR $>30.5$ dB) \\
\hline

% Mao et al. ~\cite{Mao2019_LongDistanceIMDD} &
% IMDD with OFDM-QAM (256-QAM) &
% 2.5 Gbps &
% 1000 km; Q-factor $\sim$1.6 dB lower than unencrypted OFDM-QPSK; ASE noise masking \\
% \hline

Lei \emph{et al.} ~\cite{lei202116} &
16-QAM-based QNSC with DFT-OFDM &
40 Gbps &
300 km transmission without intermediate amplifiers; DFT-OFDM reduces RX DSP complexity and OFDM PAPR; BER improvement larger than 0.00756 using optimized DSP and FCM clustering. \\
\hline

Chen \emph{et al.} ~\cite{chen2021experimental} &
Ultra-high-order ($2^{32}$) QAM &
160 Gbps (22-GBaud) &
320 km transmission; Higher-order QAM using a two-segment (SiPh) I/Q modulator driven by DACs; PDM doubles the data rate; BER below $3.9\times10^{-3}$ is achieved at OSNRs of 23 dB (regular 16-QAM) and 24 dB (encrypted 16-QAM) \\
\hline

Zhu \emph{et al.} ~\cite{zhu2023quantum} &
TQAM with secret probabilistic shaping (SPS) &
34.28 Gbps &
140 km single-span SMF; OSNR gains of 0.6 dB (TQAM/QNSC) and 1.2 dB (SPS-TQAM/QNSC) over QAM/QNSC error-free Tx ($T=372$). SPS-TQAM encrypted to $2^{10}\times2^{10}$; at Rx, BER $=7.79\times10^{-4}$ with KER = 0.0974 before information reconciliation(IR); KER reduced to 0 after Cascade iterations. \\
\hline

Zhang \emph{et al.} ~\cite{zhang2024quantum} &
3D-mapped 3-channel 64QAM with DFTs-OFDM &
80Gbps (10 GSa/s $\times$ 8 bit/symbol) &
Enhanced security via MED, NMF, and DFP; OSNR at BER $=3.8\times10^{-3}$  is 9 dB (in B2B) and 9.5 dB (160km SSMF); MED increased from 0.2857 (2D QAM) to 0.6667 (3D mapping); at OSNR = 10, NMF is 19.52 (64-3C-QAM/QNSC), DFP reaches 0.99982 at OSNR = 6; 64-3C-QAM/QNSC shows $\sim$5 dB OSNR improvement over conventional 64-QAM/QNSC. \\
\hline

Xie \emph{et al.} ~\cite{xie2025implementation} &
Deep-learning-assisted QAM &
400 Gbps per channel (21-channel WDM, total 8.4 Tbps) & 
1520 km transmission; High-order (65536) QAM with quantum noise masking and neural network-based encryption/decryption; DFP is 0.999908 (E2E-QNSC) and 0.999931 (traditional QNSC) at RX power of $-2$ dBm \\

\hline
Yuang Li \emph{et al.} ~\cite{li2023analysis} &
single-carrier PDM-16-QAM &
205.9 Gbps &
640 km transmission; OSNR penalty of 1.81 dB, including an EP of 0.67 dB; Estimated NMS $\approx$ 3.53 \& experimental/practical NMS $\approx$ 193.34; DFP $\sim$ 99.79\%; pilot-aided phase recovery reduced EP and improves linewidth tolerance. \\
\hline

Jihui Sun \emph{et al.} ~\cite{sun2023experimental} &
Probabilistically shaped QAM  &
201.6 Gbps &
Transmission over 1200 km under 20\% SD-FEC; 520 km transmission under 7\% HD-FEC; Net data rate of approximately 160 Gbps after FEC and pilot overhead; Probabilistic shaping improved BER performance and transmission reach compared with uniform QAM. \\
\hline

Dongxu Zhu \emph{et al.} ~\cite{zhu2025odd} &
Odd-order QAM &
2.94 Gbps &
25 km transmission; BER performance improved by 0.7 dB compared with random mode distribution, and Key BER equals 0 for authorized receiver and increases to 0.49 under key misalignment; DFP approaches 1 for illegal reception. 

\label{tab:IMDD_PSK_QAM}
\end{longtable}

% Abbreviations list

\section*{\textbf{Abbreviations:}} DP-PSK, Dual-polarization PSK; OSNR, optical signal-to-noise ratio; DD-MZM, dual-drive Mach–Zehnder modulator; CTFM, coarse-to-fine modulation; PDM, polarization-division multiplexing; FPGA, field-programmable gate array; TX, transmitter; RX, receiver; DSP, digital signal processing; OPLL, optical phase-locked loop; FEC, forward error correction; OFDM, orthogonal frequency division multiplexing; DFT-OFDM, discrete Fourier transform OFDM; PAPR, Peak-to-Average Power Ratio; FCM, fuzzy C-means clustering; DAC, digital-to-analog converter; SiPh, silicon photonics; TQAM, triangular quadrature amplitude modulation; SPS, secret probabilistic shaping; SMF, single-mode fiber; KER, key error rate; IR, information reconciliation; MED, minimum Euclidean distance; DFP, detection failure probability; B2B, back-to-back; SSMF, standard single-mode fiber; E2E, end-to-end; STM-16/OC-48, Y-00 Transceiver; 3C, 3 Channel; NMF, noise masking factor; NMS, number of masked signals; EP, encryption penalty; HD-FEC, hard-decision forward error correction; SD-FEC, soft-decision forward error correction; SER, symbol error rate; Q-penalty, Q-factor penalty; Tbit/s, $\mathrm{SER}_{\mathrm{eve}}$, eavesdropper symbol error rate.

% ------------------------------------------------
% section- QKD-enabled QNSC
% Image of QKD-enabled QNSC architecture

%-----------------------------------------------------
\section{Security Framework}
\label{section:5 security}

\begin{figure}[t]
\centering
\includegraphics[width=0.55\textwidth]{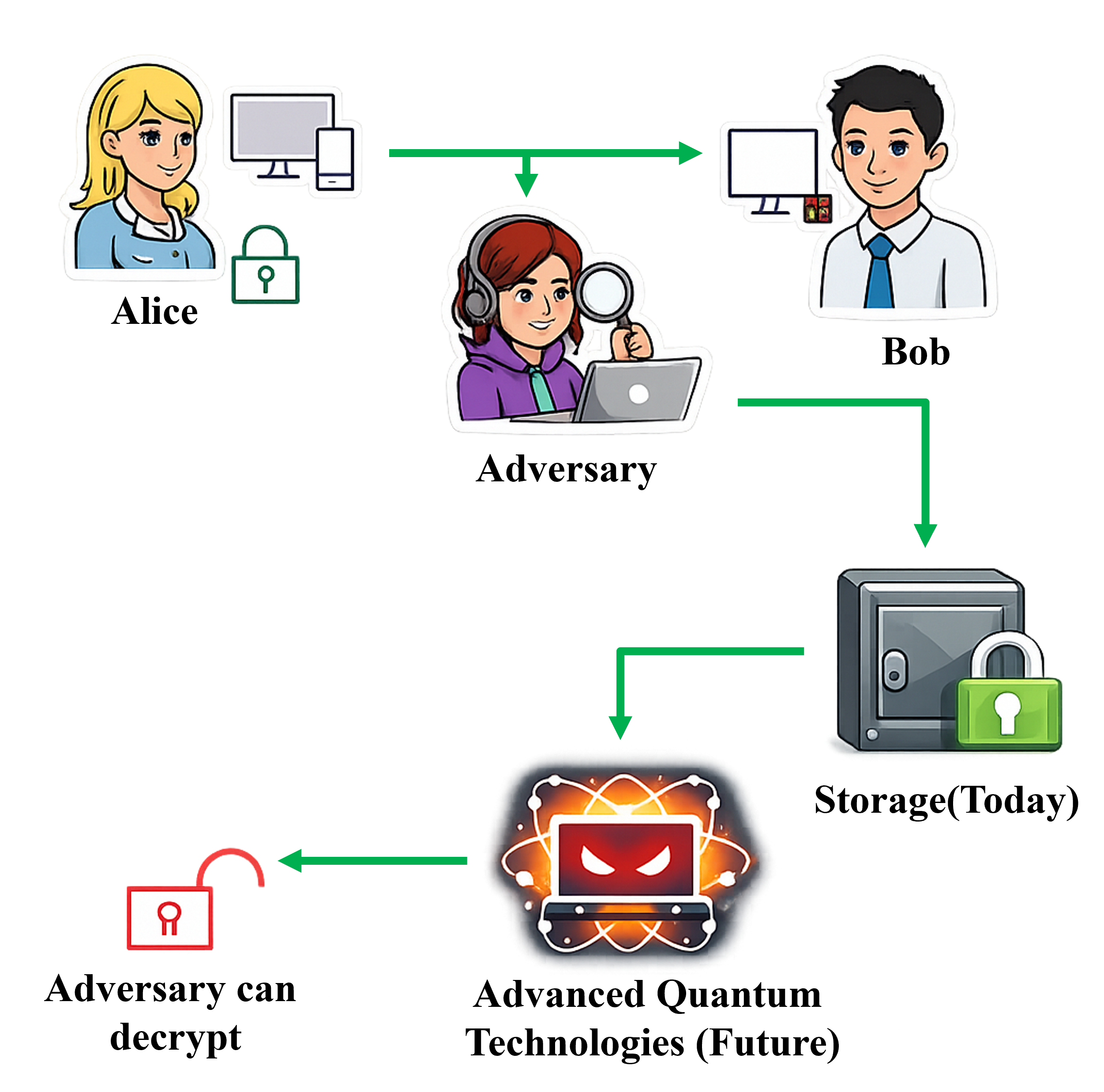}
\caption{Figure shows "Store Now, Decrypt Later", a threat from quantum storage-based attacks.   }
\Description{}
\label{fig:securityofqkd}
\end{figure}
%It is very difficult to realize an information theoretically secure symmetric key cipher based only on a mathematical algorithm. Yuen, however, pointed out that it may be possible when one employs randomization by quantum noise.
The security of the Y-00 protocol has been vigorously contested through a series of published comments, replies, and counter-replies, all of which were successfully refuted by Yuen et al. and collaborating groups through rigorous quantitative analysis ~\cite{nishioka2004much, yuen2005comment,nishioka2005reply,nair2005reply,yuen2005security,PhysRevLett.94.048902}. %Following these initial formulations, systematic cryptanalytic studies emerged, highlighting potential vulnerabilities of Y-00 under fast correlation and heterodyne-assisted attacks when linear feedback shift register (LFSR)-based key generators are employed ~\cite{donnet2006security}. 
%As Y-00 systems evolved toward higher-order modulation formats and long-haul transmission, additional attacks such as polarity inversion were identified and addressed through enhanced physical-layer countermeasures ~\cite{iwakoshi2013polarity}. More recent studies have continued this trend by analyzing correlation attacks in high-order QAM-modulated quantum noise stream ciphers, reinforcing the importance of modulation-aware security evaluation in modern Y-00 implementations ~\cite{zhang2022security}. 
The security of the Y-00 quantum noise stream cipher is rooted in fundamental limitations imposed by quantum mechanics on the processing of classical information encoded in quantum states. Several results from quantum information theory are relevant in this context; among them, three are central to understanding the physical origin of security in Y-00–type schemes. \\
The first theorem is that non-orthogonal quantum states cannot be perfectly distinguished ~\cite{helstrom1969quantum}. When logical symbols are encoded into non-orthogonal states, any measurement designed to discriminate between them is necessarily subject to a nonzero minimum error probability, even when optimal measurement strategies are employed. For example, if we prepare a state $\big|\psi\rangle$ with probability $p$ and another non-orthogonal state $\big|\phi\rangle$ with probability $1-p$, the accessible information associated with this ensemble is strictly less than the classical entropy $H(p)$. As a consequence, no measurement can allow a receiver to identify the prepared state with full reliability. A complementary limitation is provided by the second theorem, the no-cloning theorem, which states that non-orthogonal quantum states cannot be copied perfectly ~\cite{wootters2009no}. This prevents an adversary from creating multiple identical replicas of the transmitted signals, thereby improving discrimination performance through repeated or collective measurements. While this principle forms the basis for security in protocols such as BB84, it is not explicitly required for the operation of Y-00. Nevertheless, it restricts a broad class of copy-based and collective attack strategies and therefore supports the overall security framework of QNSC. According to the third theorem, the only coherent state is the input state that allows the pure state to pass through the energy loss channel. Thus, the desired state is coherent. The Y-00 protocol ensures that coherent state communication meets two criteria for quantum communication: efficiency and security, which are required for quantum cryptography. \\
The Y-00 protocol can be employed for two distinct cryptographic tasks: \emph{direct data encryption} and \emph{key generation}. These two applications are governed by different security criteria and must therefore be analyzed separately.

The principle of security underlying Y-00 is that the origin of security lies in the difference between the optimal quantum measurement performances \emph{with} and \emph{without} knowledge of the secret key. A legitimate receiver, who possesses the correct key, can perform a measurement matched to the transmitted signal states and achieve a low error probability. In contrast, an eavesdropper without access to the key is forced to measure a statistical mixture of non-orthogonal quantum states, for which quantum detection theory imposes a strictly higher minimum error probability. This asymmetry in optimal measurement performance constitutes the physical basis of security in the Y-00 protocol. The most powerful attacks on conventional stream ciphers, such as the algebraic attack and the fast-correlation attacks, also known as the known-plaintext attacks on the conventional stream ciphers, do not work on the Y-00 protocol with appropriate design, even when the key length is relatively short, i.e., of the order of $|K_s| \approx  100$. In this section, we discuss several significant attack strategies against which the security of the Y-00 protocol has been extensively analyzed in the literature.

\subsection{\textbf{Ciphertext-Only Attacks}}
In the Y-00 protocol, while ciphertext-only attacks (COA) on the data are trivially ineffective, the relevant security concern is a ciphertext-only attack on the key. Unlike conventional stream ciphers, Eve can attempt to infer the running keystream directly from measurements of the transmitted $M-$ ary quantum signals, seemingly suggesting a weakness. However, this intuition is misleading. When the running key is generated using a filtering-type PRNG consisting of an LFSR combined with a nonlinear Boolean function, the standard algebraic attacks used in classical cryptanalysis fail ~\cite{courtois2003algebraic, hirota2007practical}. In particular, recovering the initial LFSR state requires solving an overdefined system of multivariate polynomial equations derived from noisy keystream observations. Due to unavoidable quantum measurement noise, Eve’s observed keystream bits contain errors, leading to an exponential number of candidate equation systems consistent with the observations. Even with optimal Gröbner-basis-based solvers, the attacker faces a multiplicity of solutions scaling as $\mathcal{N}_{eq} = \mathcal{Q}_1=\Gamma (|\kappa|)^{|K_s|/\log M}$ ~\cite{hirota2007practical}. Collecting many measured bits $\mathcal{N} > |K_s|$ as the running key to set the overdefined equations. Then the possible number of equations for the overdefined system is $ \mathcal{N}_{eq}= \mathcal{Q}_2= \Gamma(|\kappa|)^{N/\log{M}}$ with no reliable method to identify the correct key. Consequently, ciphertext-only algebraic attacks on the Y-00 protocol become computationally infeasible, demonstrating that direct access to noisy keystream information does not compromise key security ~\cite{hirota2007practical,nair2007security}.

\subsection{\textbf{Known-Plaintext Attacks}}
In known-plaintext attacks, Eve has access to certain plaintexts along with their corresponding ciphertexts. In this scenario, modifying the known plaintext lessens the error region of the running key. Nevertheless, when employing overlapping shift keying (OSK), this region becomes equivalent to that found in a ciphertext-only attack. Eve is able to derive the running keystream irrespective of the order of the plaintext. In traditional cryptography, the algebraic attack pertains to analyzing the running key itself, which is also characterized as the known-plaintext attack ~\cite{courtois2003fast,courtois2003algebraic, hirota2007practical}. This indicates that having knowledge of the plaintext does not benefit the algebraic attack on the Y00 protocol. Eve still has the alternative of $\mathcal{Q}_1$ after the computation of the Grobner bases. However, in the context of the known plaintext attack, she can resort to a brute-force search. In other words, she has the ability to match the known plaintext against the outcome of the binary detection process using the potential key. Thus, in theory, she can decrypt the Y-00 protocol. Nevertheless, the complexity is given by $\mathcal{Q}_2 O(|K_s|^{dw})\mathcal{Q}_1 >> 2^{|K_s|}$, where $O(|K_s|^{dw})$ refers to the complexity associated with the non-linear filtering itself. For the Toyocrypt cipher, this complexity is relatively low, yet it necessitates the computation of the Gröbner bases complexity $\mathcal{Q}_2$ times. The security feature mentioned above cannot be achieved by non-random ciphers in traditional cryptography ~\cite{Hirota2005_Y00Design, hirota2007practical}.
\subsection{Correlation and Fast-Correlation Attacks}
Correlation attacks, introduced by Siegenthaler~\cite{siegenthaler1985decrypting}, exploit 
a statistical bias between the keystream of a PRNG and its constituent LFSRs. By treating 
each register independently, an attacker searches over $2^{L_i}$ candidate states for 
register $i$, reducing complexity from $\prod_i 2^{L_i}$ to $\sum_i 2^{L_i}$. The fast 
correlation attack of Meier and Staffelbach~\cite{forre1989fast} further improves this by 
recasting the problem in coding-theoretic terms: the LFSR output is treated as a codeword 
transmitted over a binary symmetric channel (BSC) with crossover probability $p_e$, and 
the initial state is recovered via iterative probabilistic decoding using low-weight 
parity-check equations derived from the feedback polynomial $c(X)$—though this requires 
fewer than roughly ten feedback taps~\cite{forre1989fast}.

In Y-00, Eve intercepts a running key $K_I$ under a known-plaintext model. Quantum and 
amplifier noise corrupt $K_I$ so that each bit agrees with the true running key $K_R$ 
with probability $p = 1 - P_m$. Given PRNG correlation $q = \Pr(K_R = x_n)$, the 
effective crossover probability available to Eve is
\begin{equation}
    p_e = 1 - (p + q) + 2pq.
\end{equation}
Both attack variants require $p_e$ to deviate sufficiently from $\frac{1}{2}$. Y-00 
defeats them through two mechanisms. First, quantum shot noise from coherent-state 
detection is irreducible: as modulation order $M$ increases, $P_m$ grows and drives 
$p \rightarrow \frac{1}{2}$, forcing $p_e \rightarrow \frac{1}{2}$ regardless of $q$. 
Once $p_e$ is sufficiently close to $\frac{1}{2}$, the correlation metric
\begin{equation}
    \alpha = L - 2\sum_{n=1}^{L}\left(K_{I,n} \oplus x_n\right)
\end{equation}
becomes statistically indistinguishable under correct and incorrect key hypotheses, causing 
the hypothesis test to fail~\cite{donnet2006security}. Second, employing a nonlinear 
filter generator of algebraic degree $d$ raises the effective linear complexity to
\begin{equation}
    \mathcal{C}_L = \sum_{i=1}^{d} \binom{|K_s|}{i},
\end{equation}
which grows exponentially in $|K_s|$~\cite{mihaljevic2005cryptanalysis}, eliminating the 
sparse polynomial structure the fast correlation attack requires. Moreover, Eve's noise is 
not a classical BSC---it is state-dependent quantum measurement noise with statistics tied 
to the coherent-state detection process, invalidating the linear code framework underlying 
the fast variant~\cite{mihaljevic2005cryptanalysis, donnet2006security}. Residual risk is 
further mitigated by periodic seed-key refresh and by keeping optical power low enough to 
deny Eve a sufficient SNR for reliable symbol discrimination~\cite{zhang2022security}.

\subsection{Polarity-Inversion Physical Attack}

A polarity inversion attack introduced by ~\cite{iwakoshi2013polarity} is an active attack in which an adversary intentionally modifies 
the transmitted signal to falsify the message received by the legitimate user. In optical and 
quantum communication systems, where information is encoded in physical parameters such as 
intensity, phase, or polarization, this attack works by transforming a signal associated with 
one logical value (e.g., $0$) into that of the opposite value (e.g., $1$). In conventional 
cryptographic systems and many QKD schemes such as BB84, protection against such attacks 
relies on classical authentication at the public communication layer rather than on any 
physical-layer mechanism; without it, polarity inversion can lead to deterministic message 
falsification, particularly under known-plaintext conditions.

The situation differs fundamentally in the Y-00 ($\alpha\eta$) quantum noise stream cipher. 
Alice encodes each plaintext bit into a coherent optical signal whose level is determined 
jointly by the plaintext and a running key shared with Bob. To execute a polarity inversion, 
an adversary must estimate the transmitted signal level and apply a precise physical shift to 
the conjugate level corresponding to the opposite logical value. This estimation is 
fundamentally limited by quantum noise in coherent-state measurement: without knowledge of 
the running key, the adversary cannot identify the correct inversion operation and risks 
applying an incorrect transformation. Consequently, even under strong assumptions such as a 
known-plaintext attack, the probability of successful symbol falsification is strictly less 
than unity.

Resistance is further strengthened by extended modulation schemes in which signal levels are 
grouped into larger communication bases, multiplying the number of possible inversion choices 
per symbol and reducing the likelihood that any single attempt succeeds. More broadly, 
polarity inversion attacks illustrate a distinctive feature of quantum noise--based 
cryptography: physical-layer uncertainty protects not only confidentiality but also message 
integrity, allowing Y-00 protocols to inherently suppress certain classes of active attacks 
without relying solely on classical authentication.
\subsection{Collective Attacks}
In the security analysis of quantum cryptographic protocols, \emph{collective attacks} 
represent an important intermediate eavesdropping strategy, positioned between individual 
attacks—in which Eve measures each signal immediately after interception—and fully 
coherent attacks, in which Eve interacts jointly with all signals simultaneously. In a 
collective attack, Eve applies the same fixed unitary coupling to each transmitted signal 
independently, attaching a probe in a fixed initial state and retaining it in quantum 
memory~\cite{biham2002security, navascues2005security}. She then performs a joint 
\emph{collective measurement} on the accumulated probe state
\begin{equation}
    \rho_E^{(n)} = \bigotimes_{i=1}^{n} \rho_E^{(x_i)}
\end{equation}
after observing $n$ signals, exploiting all available quantum correlations before 
committing to a measurement strategy.

The maximum classical information Eve can extract is bounded by the \emph{Holevo 
quantity}~\cite{furrer2012continuous}
\begin{equation}
    \chi(A:E) = S(\rho_E) - \sum_x p_x\, S\!\left(\rho_E^{(x)}\right),
\end{equation}
where $S(\rho) = -\mathrm{Tr}(\rho \log \rho)$ is the von Neumann entropy, 
$\rho_E = \sum_x p_x\, \rho_E^{(x)}$ is Eve's average probe state, and $\{p_x\}$ are 
the signal probabilities. A positive secret information rate is achievable whenever
\begin{equation}
    R_{\mathrm{sec}} = I(A:B) - \chi(A:E) > 0,
\end{equation}
where $I(A:B)$ is the mutual information between Alice and Bob. Collective attacks are 
a widely adopted threat model because they capture realistic adversarial capabilities: 
Eve requires quantum memory and optimized joint measurements, but no coherent 
multi-signal interaction.

In Y-00 quantum noise stream ciphers (QNSCs), Alice transmits coherent states encoding 
both data and running key. Under a collective attack, Eve intercepts and stores each 
coherent state before performing an optimal collective measurement to distinguish among 
non-orthogonal signal states. Security is evaluated by computing Eve's Holevo information 
separately for data and key, yielding a system-level secure rate
\begin{equation}
    R = \min\bigl\{R_{\mathrm{data}},\, R_{\mathrm{key}}\bigr\}.
\end{equation}
Wang and Zhang~\cite{wang2021security} showed that for $M_b = 31$ signal bases and 
transmission distances up to 300~km, a strictly positive secure rate is achievable, 
with security fundamentally limited by quantum noise and the non-orthogonality of the 
coherent signal constellation. Since adjacent coherent states cannot be perfectly 
discriminated by any measurement including an optimal collective one—an irreducible 
error floor is imposed on Eve, preserving a rate advantage for the legitimate parties.

Hybrid QKD-assisted QNSC systems exploit the complementary strengths of both approaches: 
QKD distributes short, information-theoretically secure seed keys with provable security 
against collective and coherent attacks (see Fig.~\ref{fig:securityofqkd}), which are 
then used in the QNSC to drive high-speed encryption, with security further reinforced 
by quantum noise masking and non-orthogonal signal modulation.

\section{QKD-integrated QNSC}
\label{section:4 QKD+QNSC}
\begin{figure}[t]
\centering
\includegraphics[width=\textwidth]{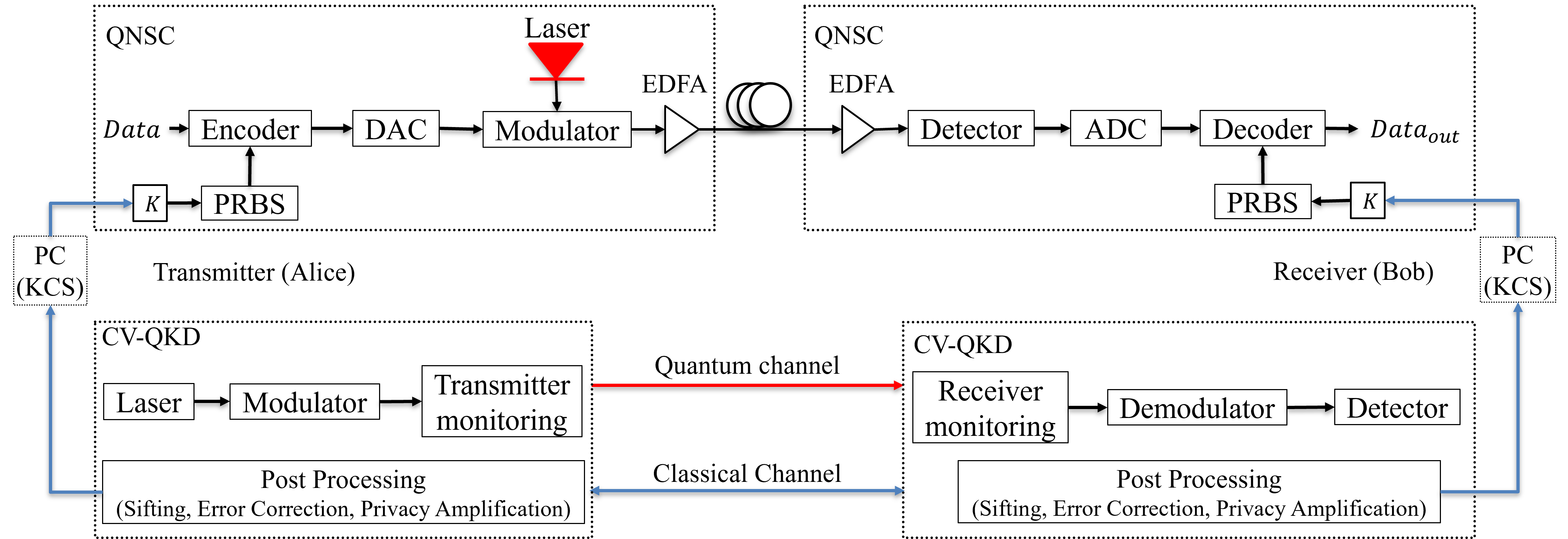}
\caption{Integrated QKD–QNSC secure optical communication architecture, where the lower layer (CV-QKD) establishes synchronized secret keys through quantum and classical channels, and the generated keys are delivered via the key control system (KCS) to the upper QNSC layer for PRBS-assisted encryption, optical fiber transmission, and receiver-side decryption.}
% \caption{QKD-enabled QNSC architecture. Secret keys generated via QKD in the lower layer and are used in basis selection/modulation in the QNSC layer, enabling high-speed secure transmission through key-dependent modulation and physical noise masking.}
\label{fig:QKD_QNSC_architecture}
\end{figure}

QNSC systems can support data rates comparable to those of classical optical communication systems. However, their long-term security relies critically on the secure generation and periodic refreshing of the seed keys that govern basis selection and modulation parameters. This requirement has motivated the integration of QKD, particularly CV-QKD, into QNSC architectures. In QKD, security is guaranteed by fundamental principles of quantum mechanics, such as the uncertainty principle, which ensures that any eavesdropping attempt inevitably introduces detectable disturbances. Once a secure key is received, it can be used to refresh the seed key of the QNSC system, thereby enhancing long-term security beyond that achievable with static pre-shared keys. This integration is illustrated in a conceptual QKD-enabled QNSC architecture in Fig.~\ref{fig:QKD_QNSC_architecture}. Here, a QKD subsystem generates fresh secret keys that are supplied to the QNSC subsystem, where transmitter and receiver use the same keys for basis selection and modulation control. Additionally, the system shows a transmission path with noise masking applied for secure high-speed communication.

Table \ref{tab:qkd+qnsc} summarizes the limited literature available for the QKD-integrated QNSC systems. Early progress toward QKD-integrated QNSC systems was demonstrated by Lu \emph{et al.}~\cite{Lu2012CSIPHomodyne}, who reported the first experimental realization of a QNSC assisted by CV-QKD. Their system employed weak coherent states and homodyne detection in an all-fiber dual unbalanced Mach--Zehnder interferometer, achieving secure communication at a modulation rate of $450~\mathrm{kb/s}$ over $27~\mathrm{km}$ of optical fiber. In this implementation, the master secret key $K_m$ was generated via a CV-QKD protocol operated using time-division multiplexing and subsequently extended to produce session keys for stream cipher encryption. The private message was encoded onto weak optical pulses whose energy was comparable to the quantum noise level, such that an unauthorized receiver experienced a severe signal-to-noise degradation, while the legitimate receiver could reliably recover the data using homodyne detection. This work established the feasibility of replacing pre-shared keys in QNSC systems with QKD-generated keys and demonstrated the compatibility of QNSC with CV-QKD infrastructures. 

\begin{longtable}{|
p{2cm}|
p{2.5cm}|
p{3 cm}|
p{1.5cm}|
p{1.2cm}|
c|}
\caption{Progress of Hybrid QKD--QNSC Systems for Secure Optical Communications.}
\label{tab:qkd+qnsc} \\

\hline
\textbf{Reference} &
\textbf{QKD Scheme} &
\textbf{QNSC Modulation format} &
\textbf{Data Rate} &
\textbf{Distance} &
\textbf{WDM} \\
\hline
\endfirsthead

\hline
\textbf{Reference} &
\textbf{QKD Scheme} &
\textbf{QNSC Modulation format} &
\textbf{Data Rate} &
\textbf{Distance} &
\textbf{WDM} \\
\hline
\endhead

\hline
\endfoot

\hline
\endlastfoot

Yuan Lu \textit{et al.} ~\cite{Lu2012CSIPHomodyne} &
Gaussian Modulation CV-QKD &
M-PSK &
450 Kbps &
27.2 km &
No \\
\hline

Nakazawa \textit{et al.} ~\cite{nakazawa2016real, nakazawa2017qam, nakazawa2018secure} &
Discrete Modulation CV-QKD &
QAM &
70 Gbps &
100 km &
No \\
\hline

Shi \textit{et al.} ~\cite{shi202210} &
QNSC-based QKD &
Spectral Phase Encoding &
10 Gbps per channel &
100 km &
Yes \\

\end{longtable}
Despite its conceptual significance, the achievable data rate in the homodyne-based weak-signal regime was inherently limited by detector bandwidth and signal power constraints. To overcome these limitations and enable quantum-secured communication at classical optical transmission speeds, subsequent efforts focused on combining QNSC with high-capacity coherent modulation formats. A major milestone in this direction was reported by Nakazawa \emph{et al.}~\cite{nakazawa2017qam}, who demonstrated the first on-line, high-speed, and large-capacity secure optical communication system based on QAM-based QNSC integrated with CV-QKD. 
In their architecture, multilevel QAM was employed, and encryption was realized through key-controlled basis randomization combined with intentional masking by quantum and ASE noise. This design ensured that, without access to the correct running key, the encrypted constellation points were completely buried within noise, rendering symbol discrimination fundamentally unreliable for an eavesdropper. Experimentally, Nakazawa \emph{et al.} demonstrated real-time secure transmission over $100~\mathrm{km}$ of standard single-mode fiber using polarization-multiplexed 128-QAM at a symbol rate of $5~\mathrm{Gsymbols/s}$. This resulted in a secure data transmission rate of $70~\mathrm{Gbit/s}$ with a spectral efficiency of $10.3~\mathrm{bits/s/Hz}$, representing one of the highest data rates reported for quantum-secured optical communication under realistic assumptions.

The CV-QKD subsystem employed QPSK-modulated weak coherent states and homodyne detection to generate secret keys at rates of approximately $600$--$700~\mathrm{bit/s}$ under a beam-splitting attack model. Although this key rate was several orders of magnitude lower than the encrypted data rate, it was sufficient to refresh the QNSC seed keys every $0.5$--$1~\mathrm{s}$ during continuous operation. The authors further quantified security using the detection failure probability (DFP), which exceeded $99.9\%$ for an unauthorized receiver over the full $100~\mathrm{km}$ transmission distance, while the legitimate receiver achieved error-free decryption after forward error correction. %These results clearly demonstrate that integrating high-order QAM-based QNSC with CV-QKD enables long-distance, high-capacity secure optical communication without being constrained by the key-rate bottleneck inherent to one-time-pad-based QKD systems. \\
In addition to QAM-based QNSC architectures, an alternative realization of QKD-assisted quantum noise encryption at the physical layer was reported by Shi and Xiao~\cite{shi202210}. In this work, the authors demonstrated a $10~\mathrm{Gb/s}$ secure optical transmission system by combining optical physical-layer encryption with QKD based on a Y--00/QNSC. The physical-layer encryption was implemented using spectral phase encoding, which induces temporal pulse spreading and overlap, while the QKD subsystem was employed to securely distribute the dynamically updated encryption parameters.
Experimentally, the scheme was validated in a $20$-channel WDM network over $100~\mathrm{km}$ of standard single-mode fiber. The QKD system operated with an encryption key update rate of up to $10~\mathrm{MHz}$, enabling frequent refreshing of the physical-layer encryption parameters. Error-free transmission ($\mathrm{BER} \leq 10^{-9}$) was achieved for all WDM channels over the full transmission distance, with a power penalty below $3~\mathrm{dB}$ introduced by the encryption process. In contrast, an unauthorized receiver experienced a BER exceeding $0.48$, independent of the received optical power, demonstrating a clear security gap between cooperative and non-cooperative users. %These results confirm that integrating QKD with quantum noise–based physical-layer encryption provides an effective and WDM-compatible approach for securing high-rate optical transmission while maintaining acceptable system performance.

% ------------------------------------------------
\section{Performance, Challenges, and Future Directions}
\label{section:6 challenges and future direction}

\subsection{Performance and Challenges}
QNSC enhanced by QKD allows for the frequent and theoretically secure updating of seed keys used in quantum noise-based stream encryption. Even with moderate QKD key rates, it is possible to refresh the seed key after using only a small portion of a PRBS period, thereby minimizing risks associated with key reuse and correlation attacks. The security of the QNSC system is further bolstered by the hybrid QKD+QNSC system, which requires an eavesdropper to breach both the QKD key-distribution channel and the QNSC data channel simultaneously, significantly increasing the difficulty and likelihood of detection of any attack. Additionally, QKD-assisted QNSC systems are highly compatible with current commercial optical communication infrastructures, making them ideal for ultrahigh-capacity secure data transmission. Recent experimental demonstrations have confirmed the practical feasibility of this method (Table \ref{tab:qkd+qnsc}). Collectively, these findings suggest that QKD-assisted QNSC offers a promising path to bridging the gap between information-theoretic security and the bandwidth demands of modern optical networks.

 Despite these promising performance gains, several challenges remain for large-scale deployment. Practical QNSC systems are sensitive to environmental effects such as temperature fluctuations, fiber birefringence, phase drift, and polarization instability, all of which can degrade constellation stability and quantum noise masking efficiency over long-term operation. In addition, synchronization between the QKD subsystem and the high-speed QNSC data channel introduces system-level complexity, particularly under realistic network conditions involving fiber loss, excess noise, and finite-key effects. From a security standpoint, while QKD-assisted key renewal significantly mitigates correlation and key-reuse attacks, rigorous security analyses under realistic noise models and advanced attack strategies remain an open problem, especially for high-order modulation formats.

\subsection{Future Directions}
To advance QKD-assisted QNSC systems toward large-scale use, several key developments are necessary. While current implementations using CV-QKD have proven effective for high-speed communication within metropolitan areas, future efforts should focus on increasing transmission distances. This can be achieved by integrating cascaded or trusted-node QKD architectures, which allow for the repeated delivery of quantum keys. Such frameworks would enable secure, long-haul fiber networks to maintain periodic seed-key updates over hundreds of kilometers. Beyond fiber-based links, FSO implementations of QNSC offer an attractive platform for short- to medium-range secure communication. Experimental demonstrations have shown error-free Y-00/QNSC transmission at data rates of several gigabits per second over short-range FSO links, with secure transmission verified using high-order IM formats over tens of meters. 
Recent research suggests that FSO links using QNSC can operate over several kilometers, provided there is adequate signal amplification and a high signal-to-noise ratio. However, current modeling has moved beyond simple BER analysis. Because atmospheric turbulence can degrade optical signals, researchers are now focusing on waveform reconstruction fidelity—how accurately the receiver can recreate the original signal, and statistical indistinguishability, which ensures the encrypted signal cannot be distinguished from random noise by an attacker. This shift provides a more comprehensive way to evaluate system security and performance in unpredictable environments. Another potential arena of research involves applying QNSC to visible light communication (VLC) and UOWC. Secure communication in marine environments is traditionally difficult because water causes significant signal loss (attenuation), scattering, and high levels of background noise. Because QNSC is inherently designed to operate within noisy environments, it is uniquely suited for underwater use. When paired with quantum-assisted key management, QNSC could provide a robust framework for secure, short-range communication in naval operations and subsea sensor networks where classical encryption methods may struggle. From a protocol standpoint, developing hybrid architectures that combine DV-QKD with QNSC is a vital next step. While QNSC provides high-speed data encryption at the physical layer, integrating it with DV-QKD can offer stronger security guarantees (tighter security bounds). Lastly, a key open challenge is the development of rigorous security proofs that account for collective and coherent attacks under real-world conditions. While QKD-assisted key renewal significantly strengthens resistance against correlation and key-reuse attacks, extending security proofs to encompass advanced quantum attack models like collective ~\cite{wang2021security,navascues2005security,furrer2012continuous} and Trojan-horse attacks ~\cite{ma2016quantum, pan2020practical} remains essential for establishing long-term confidence in QNSC-based systems. The integration of deep learning with QNSC opens promising avenues for next-generation secure optical communication systems. Machine-learning-based encoders and decoders can be further explored to adaptively optimize constellation shaping, encryption, and decoding under varying channel conditions, particularly in the presence of fiber nonlinearities and hardware impairments. Recent demonstrations of ultra-high-rate QNSC transmission indicate that such learning-assisted frameworks can support secure communication at backbone-network scales ~\cite{xie2025implementation}. Extending these approaches to multi-channel, dynamic, and heterogeneous network scenarios remains an important avenue for future investigation.

\section{Conclusion}
QKD and QNSC have evolved in parallel, each addressing different aspects of secure communication and each carrying its own strengths and limitations. QNSC enhances data security directly at the physical layer by exploiting quantum noise as a protective resource. When carefully designed, this noise-induced masking makes the ciphertext intrinsically difficult to analyze, offering strong resistance to ciphertext-only and correlation-based attacks.
QKD, on the other hand, does not encrypt data itself but provides a provably secure method for distributing cryptographic keys. Security has been thoroughly examined for several well-known protocols, and it has been shown, at least in theory, that these protocols can withstand strong adversaries capable of carrying out coherent or collective quantum attacks.
QKD, when integrated with QNSC, provides an advanced quantum-secure technology in which the secure keys generated by the QKD protocol are used to generate a running key in the QNSC setup, thereby enabling secure data encryption. 
This work has presented a unified perspective on QNSC, covering its operating principle, various modulation schemes, and its security against various attacks. We have further highlighted the experimental progress demonstrating multi-gigabit to terabit secure transmission over fiber and FSO links, confirming the practical feasibility of quantum noise masking encryption in practical scenarios. Despite these advantages, important challenges like environmental instabilities and synchronization between QKD and high-speed data channels still remain. %Further, the extension of QNSC to free-space, visible-light, and underwater optical platforms introduces new performance metrics beyond conventional BER, including waveform reconstruction fidelity and statistical indistinguishability. The emergence of learning-assisted constellation optimization and multidimensional encryption further broadens the design space for next-generation secure optical networks. 
QKD-integrated QNSC thus offers a promising route towards bridging the gap between information-theoretic security and the bandwidth demands of the modern communication infrastructure. By combining quantum secure key with quantum noise based physical layer encryption architecture provides a scalable and technologically compatible framework for future high-capacity secure networks. Continued progress in security proofs, experimental stabilization, and system-level integration will be essential to fully realize the long-term potential of quantum-noise–based cryptographic systems.
%%
%% The acknowledgments section is defined using the "acks" environment
%% (and NOT an unnumbered section). This ensures the proper
%% identification of the section in the article metadata, and the
%% consistent spelling of the heading.
\begin{acks}
This work has been partially funded by the project grants US-India Science and Technology Endowment Fund \# USISTEF/QT/160/2023 and by the Department of Science and Technology, India, under National Quantum Mission \# NQM/QC/4262.
\end{acks}

%%
%% The next two lines define the bibliography style to be used, and
%% the bibliography file.
% \bibliographystyle{unsrt}
\bibliographystyle{ACM-Reference-Format}
\bibliography{my_references}

%%
%% If your work has an appendix, this is the place to put it.
%\appendix

%\section{Research Methods}

\end{document}